\documentclass[12pt]{article}
\usepackage{epsfig}

\usepackage{a4}

\usepackage{amsmath}

\begin{document}

\title{ The effect of Chern-Simons dynamics 
on the energy of electrically charged and spinning
vortices  
}
\author{{\large Francisco Navarro-L\'erida,}$^{1}$
{\large Eugen Radu}$^{2}$
and {\large D. H. Tchrakian}$^{3,4}$ 
\\ 
\\
$^{1}${\small Dept. de F\'isica At\'omica, Molecular y Nuclear, Ciencias F\'isicas,}\\
{\small Universidad Complutense de Madrid, E-28040 Madrid, Spain}\\  
$^{2}${\small Departamento de Fisica da Universidade de Aveiro and }\\
{\small Center for Research and Development in Mathematics and Applications (CIDMA),}
\\
 {\small 
 Campus de Santiago, 3810-183 Aveiro, Portugal
}\\ 
$^{3}${\small School of Theoretical Physics, Dublin Institute for Advanced Studies,}
\\{\small 10 Burlington Road, Dublin 4, Ireland }
\\
  $^{4}${\small  Department of Computer Science, NUI Maynooth, Maynooth, Ireland}}

\date{\today}
\newcommand{\dd}{\mbox{d}}
\newcommand{\tr}{\mbox{tr}}
\newcommand{\la}{\lambda}
\newcommand{\vr}{\varrho}
\newcommand{\ka}{\kappa}
\newcommand{\f}{\phi}
\newcommand{\vf}{\varphi}
\newcommand{\F}{\Phi}
\newcommand{\ta}{\theta}
\newcommand{\Ta}{\Theta}
\newcommand{\al}{\alpha}
\newcommand{\bt}{\beta}
\newcommand{\ga}{\gamma}
\newcommand{\de}{\delta}
\newcommand{\si}{\sigma}
\newcommand{\Si}{\Sigma}
\newcommand{\bnabla}{\mbox{\boldmath $\nabla$}}
\newcommand{\bomega}{\mbox{\boldmath $\omega$}}
\newcommand{\bOmega}{\mbox{\boldmath $\Omega$}}
\newcommand{\bsi}{\mbox{\boldmath $\sigma$}}
\newcommand{\bchi}{\mbox{\boldmath $\chi$}}
\newcommand{\bal}{\mbox{\boldmath $\alpha$}}
\newcommand{\bpsi}{\mbox{\boldmath $\psi$}}
\newcommand{\brho}{\mbox{\boldmath $\varrho$}}
\newcommand{\beps}{\mbox{\boldmath $\varepsilon$}}
\newcommand{\bxi}{\mbox{\boldmath $\xi$}}
\newcommand{\bbeta}{\mbox{\boldmath $\beta$}}
\newcommand{\ee}{\end{equation}}
\newcommand{\eea}{\end{eqnarray}}
\newcommand{\be}{\begin{equation}}
\newcommand{\bea}{\begin{eqnarray}}

\newcommand{\ii}{\mbox{i}}
\newcommand{\e}{\mbox{e}}
\newcommand{\pa}{\partial}
\newcommand{\Om}{\Omega}
\newcommand{\om}{\omega}
\newcommand{\vep}{\varepsilon}
\newcommand{\bfph}{{\bf \phi}}
\newcommand{\lm}{\lambda}
\def\theequation{\arabic{equation}}
\renewcommand{\thefootnote}{\fnsymbol{footnote}}
\newcommand{\re}[1]{(\ref{#1})}
\newcommand{\R}{{\rm I \hspace{-0.52ex} R}}
\newcommand{\N}{{\sf N\hspace*{-1.0ex}\rule{0.15ex}%
{1.3ex}\hspace*{1.0ex}}}
\newcommand{\Q}{{\sf Q\hspace*{-1.1ex}\rule{0.15ex}%
{1.5ex}\hspace*{1.1ex}}}
\newcommand{\C}{{\sf C\hspace*{-0.9ex}\rule{0.15ex}%
{1.3ex}\hspace*{0.9ex}}}
\newcommand{\eins}{1\hspace{-0.56ex}{\rm I}}
\renewcommand{\thefootnote}{\arabic{footnote}}

\maketitle


\bigskip

\begin{abstract}
We study the effect of a Chern-Simons term  on the electrically charged and spinning solitons of several $U(1)$ gauged models in $2+1$
dimensions. These are vortices of   complex scalar field theories, both with and without symmetry breaking dynamics, and
  the $O(3)$ Skyrme model. 
In all cases  the gauge decoupling limits are also considered. 
It is well known that the effect of the Chern-Simons dynamics is to endow vortices with electric charge $Q_e$ and spin
$J$, but our main aim here is to reveal a new feature: 
that the mass-energy $E$ of the electrically charged vortex can be lower than that of the electrically neutral one,
in contrast to the usual monotonic increase of $E$ with $Q_e$. These effects of Chern-Simons dynamics were observed previously in
$3+1$ dimensional systems, and the present results can be viewed as corroborating the latter.
Moreover, the usual energy-spin relationship
is likewise altered. 
We carry out a detailed quantitative analysis of azimuthally symmetric vortices and describe their qualitative features by constructing the
solutions numerically. 
 
\end{abstract}
\medskip
\medskip

\newpage

\section{Introduction}

The study of  electrically charged and spinning solutions of $U(1)$ gauged models in $2+1$ dimensions
can be traced back at least to the work of Julia and Zee \cite{Julia:1975ff}.
As shown there, the Nielsen-Olesen vortices
\cite{Nielsen:1973cs} (which are solutions of the Abelian gauged Higgs model)
do not possess spinning  (and electrically charged)  generalisations with finite energy.
This feature can be attributed to the long range behaviour of the electric field, whose effect is
present also for $U(1)$ gauged models without a symmetry breaking scalar field.

A standard way to circumvent this obstacle is to add a Chern-Simons (CS) term to the gauge-field Lagrangian.
Chern-Simons field theory in $2+1$ dimensions has featured prominently in the literature since the seminal work of Refs.~\cite{Deser:1982vy,Deser:1981wh}.
The salient effect introduced by the CS dynamics is the endowment of electric charge and angular momentum to the solitons,
while preserving a finite mass. The novel effect of the CS dynamics revealed in the present work is, 
that the mass of the electrically charged soliton
can be smaller than the mass of its neutral counterpart, and the usual energy-spin relationship can likewise be altered.

The nature of this mechanism peculiar to Chern-Simons (CS) dynamics, is quite subtle.
Clearly the value of the electric charge depends on the strength of the CS coupling, say $\kappa$, but if the
dependence of the mass/energy is tracked by varying $\kappa$ then the $usual$ monotonic increase of the energy is observed.
The crucial feature of the CS dynamics is that the static solutions are characterised by a free paramereter
$b_{\infty}$, which is
the asymptotic value of the function describing electric component $A_0$ of the Abelian field. (Note that it
is the presence of $\kappa$
in the Lagrangian which renders $A_0$ nontrivial\footnote{A subtlety here is that for some value of $b_{\infty}$ for which $A_0\neq 0$,
the electric charge may vanish (with higher electric multipoles being present).}.)
It is the dependence of the electric charge (and the angular momentum)
and the mass/energy on $b_{\infty}$ that enables the mutual tracking of these quantities. In some models where $b_{\infty}$
is not a free parameter of the solutions, this mechanism is absent and the dependence of the energy on the electric charge can only be tracked by varying $\ka$, which amounts to
charging the theory. In those cases, the energy increases monotonically with the electric charge (and the angular momentum). This monotonic behaviour is present also in a theory allowing for a free value of
$b_{\infty}$, when the latter is held fixed and $\ka$ in increased.

This analysis is carried out (broadly) for three types of models supporting finite energy static solutions in $2+1$ dimensions.
These are: 
${\bf i)}$ the $U(1)$ gauged complex scalar field model with a scalar field vanishing in the far field,
${\bf ii)}$ 
the $U(1)$ gauged complex scalar field model with symmetry breaking dynamics,
and  ${\bf iii)}$ the $U(1)$ gauged  $O(3)$ Skyrme model.

The first one of the complex scalar models features no symmetry breaking,
 supporting, however, nontopological vortices
with finite mass, angular momentum and electric charge ~\footnote{The gauge decoupled
version of these supporting $Q$-vortices is also considered in passing.}. The second class of complex scalar models, which exhibits symmetry breaking dynamics, supports Abelian Higgs vortices that are
topologically stable $prior$ to the introduction of the Chern-Simons (CS) term.
In both classes of models CS terms, which are not positive definite by construction, are prominently present in the Lagrangians and provide the new features of the vortices under investigation here.
Thus the question of topological stability is not considered as the important feature. 
In the case of the Abelian Higgs systems, both the $p=1$ and the $p=2$ models are studied~\footnote{In each space dimension $D$,
 a hierarchy of $SO(D)$ gauged Higgs models can be defined, that support
finite energy topologically stable solutions (monopoles in $D\ge 3$ and vortices in $D=2$). This hierarchy, labelled by $p$, consists of models of increasing nonlinearity with increasing $p$
up to the maximum allowed $p$ for each $D$. The $p=1$ models in $D=2$ and $D=3$ respectively, are the usual Abelian Higgs and the Georgi-Glashow (in the BPS limit) models. For a description of these
models, see Ref.~\cite{Tchrakian:2010ar} and references therein.}. It turns out that
in the (usual) $p=1$ model and the pure $p=2$ model, the presence of the CS term does not result in lowering the mass of the soliton with increasing electrical charge, while in the case of a hybridised $p=2$ model the mass-energy of an electrically charged vortex can be lower than that of the neutral one.

In addition to the gauged complex scalar models, 
we have considered  $U(1)$ gauged $O(3)$ Skyrme models, augmented by the usual CS term. 
These vortices are topologically stable $prior$ to the introduction of the CS term.
In this case the mass of an electrically charged vortex can be lower than that of the neutral one.

Some of the models in this work have already been under scrutiny in the literature, although from a different direction. For example, in the case of the complex scalar field model(s)
{\bf not} featuring symmetry breaking dynamics, such vortices were described in Refs.~\cite{Khare:1989dx}, and \cite{DeshaiesJacques:2006ae}.   
  In the case of  a complex scalar field featuring symmetry breaking dynamics, $i.e.$ the Abelian Higgs (AH) model, 
the CS term was added to the Maxwell-Higgs Lagrangian in \cite{Paul:1986ix}, while in \cite{Hong:1990yh,Jackiw:1990aw}
the CS term was the sole source of the gauge field dynamics~\footnote{In the latter case the electrically charged solutions are self-dual solutions attaining the absolute minimum.}.
 Here we have considered the first two in the family of $p$-Abelian-Higgs (AH) models~\cite{Burzlaff:1994tf,Arthur:1998nh}, the $p=1$ case being the $usual$ AH model. We will see that
the $p=2$ AH model displays some very interesting properties. 

Still with Abelian gauge dynamics but with the complex scalar replaced by the $O(3)$ sigma model scalar, magnetic Abelian vortices were
constructed in \cite{Schroers:1995he}. This model is the Skyrme analogue of the Abelian Higgs model~\footnote{By Skyrme systems we mean all possible $O(D+1)$ sigma models in $D$ dimensions.},
 and like the latter does not support electric charge and spin.
Adding a Chern-Simons term to this Lagrangian results in systems that
support electrically charged, spinning magnetic vortices. This was carried out in Refs.~\cite{Ghosh:1995ze,Kimm:1995mi,Arthur:1996uu,loginov}, in analogy with the Higgs models
cited above in Refs.~\cite{Paul:1986ix,Hong:1990yh,Jackiw:1990aw}~\footnote{Again, in the absence of the Maxwell term, the energy of the resulting electrically charged
vortex attains its absolute minimum~\cite{Arthur:1996uu}.}. 


Our primary objective in this work is to investigate 
in a systematic way
the relationship between the mass, 
 electric charge and angular momentum in these models,
looking for generic features,
a subject which was not addressed in the existing literature.
%
Then we recover a number of known results, 
namely that the effect of the Chern-Simons dynamics 
is to endow electric charge and angular momentum to the solitons. 
What is completely new here
is that the relation of the electric charge and the mass is quite different 
from that of  Julia-Zee dyon solutions~\cite{Julia:1975ff} of the Georgi-Glashow model,
 in the absence of CS dynamics. 
While in the latter case the mass increases with
electric charge, here the mass decreases with the charge 
in some regions of the parameter space. 
This tendency,  namely that of the energy of the dyon not increasing uniformly with increasing electric charge, 
was observed also in $3+1$ dimensional non-Abelian Higgs models  featuring (new) 
Higgs dependent CS terms~\cite{Navarro-Lerida:2014rwa}.

Finally, we mention that
in each of the $U(1)$ gauged models studied, we considered also their gauge decoupling limits.

The paper is organised as follows. 
In the next Section we introduce the general framework for the models studied, including the gauge and scalar fields Ans\"atze. 
In Section {\bf 3} we present two models with a gauged complex scalar field.
The first theory presents no symmetry breaking dynamics, whose vortices while finite energy, are
not topologically stable. Then the Abelian Higgs model(s) are analysed, which do feature symmetry breaking dynamics, so that their vortices are
topologically stable. In Section {\bf 4} we consider the $U(1)$ gauged Skyrme model, whose vortices are also topologically stable.
Finally, in Section {\bf 5} the main results are summarised.
The paper contains also two Appendices. In Appendix {\bf A} the conserved charges are defined and in Appendix {\bf B} the gauge decoupling limits for the
three models are discussed. 
\\
\\
 {\bf Conventions and notations:}
\\
\\
The backreaction of the matter 
fields on the geometry is ignored in this work. 
Thus we consider a fixed three dimensional
Minkowski spacetime background
with a line element 
\begin{eqnarray}
 ds^2=g_{\mu\nu}dx^\mu dx^\nu=dt^2-dx^2-dy^2 ,
\end{eqnarray} 
where $t$ is the time coordinate and $x,y$
are the usual Cartesian coordinates.
The same line element expressed in cylindrical coordinates $r,\theta$ (with $r=\sqrt{x^2+y^2}$ and $\theta= {\rm arctan}(x/y) $) reads
\begin{eqnarray}
\label{Mink2}
ds^2=dt^2-dr^2-r^2d\theta^2,
\end{eqnarray} 
 where
$0\leq r<\infty$ and $0\leq \theta<2\pi$.

Note that throughout the paper, the Greek indices like $\mu,\nu$ run from $0$ to $2$ (with $x^0=t$ and $\hat x_i=x_i/r$);
Latin indices like $i,j=1,2$ label space coordinates, while 
Latin letters like $a,b$ correspond to internal group indices for scalar field multiplets  with $a=\al,3$; $b=\bt,3$; and $\al,\bt=1,2$.

%

\section{The general framework}
The Lagrangians of the models studied in this work can be expressed formally as
\begin{eqnarray}
\label{gen-action}
 {\cal L}= \gamma\,{\cal L}_{\rm AC}+ \bt\,{\cal L}_{\rm AS}+ {\cal L}_{\rm CS} \ ,
\end{eqnarray}
In \re{gen-action} the term ${\cal L}_{\rm AC}$ summarises two types of $U(1)$ gauged complex scalar models and their gauge decoupled limits; ($a$) models supporting $nontopological$ vortices, and ($b$), models supporting $topological$ vortices.
The two types of models are distinguised by their respective self-interaction potentials of the complex scalar fields, in case ($b$) featuring symmetry breaking, which are the Abelian-Higgs (AH)
 models~\footnote{${\cal L}_{\rm AH}^{(p)}$ define members of the family of Abelian-Higgs (AH) models, the $p=1$ member being the usual AH model and, the more nonlinear, $p=2$ member given below by \re{pAHa}.} ${\cal L}_{\rm AH}^{(p)}$.

The term ${\cal L}_{\rm AS}$ in \re{gen-action} defines the $U(1)$ gauged Skyrme scalar~\cite{Schroers:1995he}. 
Finally,
 the term ${\cal L}_{\rm CS}$ is the Chern-Simons (CS) density 
\begin{eqnarray}
\label{lcs}
{\cal L}_{\rm CS} =  \ka  \vep^{\lambda \mu\nu}A_\lambda F_{\mu\nu}.
\end{eqnarray}
defined in terms of the $U(1)$ gauge potential $A_\mu$ and curvature $F_{\mu\nu}$.
In Sections {\bf 3} and {\bf 4}, we shall set $\bt=0$ and $\gamma=0$ in turns. 

It should be emphasised that the use of the term $topological$ is qualified. It is meant only to distinguish those vortices from the $nontopological$ ones, but since the added Chern-Simons (CS) term \re{lcs}
is not positive definite, the description $topological$ is valid only to the vortices of the model prior to the introduction of the CS term. The exceptions to this are the vortices of the models in which the Maxwell term in the Lagrangian is suppressed,
leaving only the CS term to sustain the dynamics of the $U(1)$ field. Such vortices are studied in Refs.
\cite{Hong:1990yh,Jackiw:1990aw} for the Abelian Higgs case and in Refs. \cite{Ghosh:1995ze,Arthur:1996uu} for the Abelian Skyrme.
(In these models, the energy is minimised absolutely by a Bogomol'nyi bound.)

\subsection{Imposition of azimuthal symmetry}

Prior to specifying the dynamics of the models, we find it convenient to state the Ans\"atze on the various fields studied.
All configurations in this work are static and azimuthally symmetric $i.e.$ they do not exhibit a dependence on the coordinates $(t,\ta)$
at the  level of energy-momentum tensor (although this is not excluded at the level of the field Ansatz).   

For a vortex number $n\ge 2$, symmetric solutions may not be those with the lowest energy, less symmetric solutions
with dependence of both $(r,\theta)$ having lower mass. Such solutions do exist, see $e.g.$ the recent work
\cite{Samoilenka:2016wys}, \cite{Samoilenka:2015bsf}.
Our numerical studies here, are limited to unit vorticity ($n=1$) configurations.

\subsubsection{The $U(1)$ gauge field}
Subject to azimutal symmetry, the components of the $U(1)$ connection $A_{\mu}=(A_i,A_0)$ are
\bea
\label{Ma}
A_i&=&\left(\frac{a(r)-n}{r}\right)\,\vep_{ij}\,\hat{x}_j\ ,\quad A_0=b(r) , \label{axAi0}
\eea
where the integer $n$ is the vortex number
and   $a(r)$, $b(r)$ are
the electric and magnetic potentials, respectively.

%
The field strength tensor resulting from \re{axAi0} is
\bea
F_{ij}&=&-\frac{a'}{r}\vep_{ij}\label{Fij2}\ ,\quad F_{i0}=b'\hat x_i\label{Fij02}~.
\eea
The one dimensional density $L_{\rm CS}=r{\cal L}_{\rm CS}$ resulting from the Chern-Simons density \re{lcs} is
\be
\label{redCS}
L_{\rm CS} =2\ka [(ab'-ba')-nb']~.
\ee


\subsubsection{The complex scalar field}
Models featuring the complex scalar field $\vf$ are studied in Section {\bf 3}.
We will employ both parametrisations of the scalar field, the complex scalar $\vf$ or the real doublet $\f^{\al}$ related by
\be
\label{para}
\vf=\f^1-i\f^2\ , \quad \f^{\al}=(\f^1,\f^2)\ ,\quad \al=1,2\,,
\ee
such that the covariant derivative $D_{\mu}\vf=\pa_{\mu}\vf+i\,A_{\mu}\vf$ is expressed as
\be
D_{\mu}\f^{\al}=\pa_{\mu}\f^{\al}+A_{\mu}(\vep\f)^{\al}\label{cova}\ ,\quad(\vep\f)^{\al}=\vep^{\al\beta}\f^{\beta}\,.
\ee
Subjecting the scalar field to azimuthal symmetry, we have the Ansatz
\be
\label{h}
\f^{\al}= \eta h(r)\,n^{\al}\ ,\quad n^{\al}=\left(
\begin{array}{c}
\cos n\ta\\
\sin n \ta
\end{array}
\right) ,
\ee
where $\ta$ is the azimuthal angle, $n$ is the (integer) vortex number, and $\eta>0$.

The Ansatz \re{h} results in the components of the covariant derivative \re{cova},
\be
\label{di0f}
 D_i\f^{\al}=\eta h'\hat x_i\,n^{\al}+\eta\frac{ah}{r}(\vep\hat x)_i(\vep n)^{\al}\ ,\quad  D_0\f^{\al}=\eta bh\,(\vep n)^{\al}\,.
\ee 

\subsubsection{The $O(3)$ Skyrme scalar field}
Models featuring the $O(3)$ Skyrme scalar field $\ta^a=(\ta^{\al},\ta^3)$ are studied in Section {\bf 4}.
These are subject to the constraint
\be
\label{constr1}
|\ta^a|^2=1\,\quad a=(\al,3)\ ,\ \ {\rm with}\quad \al=1,2\, ,
\ee
with the covariant derivative $D_{\mu}\ta^a=(D_{\mu}\ta^{\al},D_{\mu}\ta^3)$ defined by the gauging prescription
\bea
\label{cov}
D_{\mu} \ta^{\alpha} &=&\pa_{\mu} \ta^{\alpha}+A_{\mu}\,(\vep\ta)^{\al}\,,\qquad D_{\mu} \ta^3 =\pa_{\mu} \ta^3\,.
\eea
Subject to azimuthal symmetry, we have the Ansatz
\be
\label{fa}
\ta^a=\left(
\begin{array}{c}
\ta^{\al}\\
\ta^3
\end{array}
\right)=
\left(
\begin{array}{c}
P(r)\,n^{\al}\\
Q(r) 
\end{array} 
\right)\ ,
\ee
with to $P^2+Q^2=1$, and $n^{\al}$ given in \re{h}.
The trigonometric parametrisation of $P$ and $Q$ in terms of the radial function $f(r)$ is
\be
\label{trig0}
P(r)=\sin f(r)\ ,\quad Q (r)=\cos f(r)\ .
\ee
The Ansatz \re{fa} results in the components of the covariant derivative \re{cov}
\bea
D_i\ta^{\al}&=&P'\hat x_i\,n^{\al}+\frac{a\,P}{r}(\vep\hat x)_i(\vep n)^{\al}\ ,\quad D_0\ta^{\al}=b\,P\,(\vep n)^{\al} \ ,\label{di0fal}\\
D_i\ta^{3}&=&Q'\,\hat x_i\ ,\qquad\qquad\qquad\qquad\quad D_0\ta^{3}=b\,Q \ . \label{di0f4}
\eea

\subsection{Numerical approach}
In the absence of closed-form solutions, we relied on numerical methods to solve the field
equations for various models in this work. 
 For most of the solutions reported here, the system of coupled differential equations,
 with appropriate boundary conditions,  was solved by using the software package COLSYS developed
by Ascher, Christiansen and Russell \cite{colsys}.
 This solver employs a colocation
method for boundary-value ordinary differential equations and a damped Newton method
of quasi-linearization. At each iteration step a linearized problem is solved by using a spline
collocation at Gaussian points. Since the Newton method works very well when the initial
approximate solution is close to the true solution, the full spectra of solutions for varying
various parameters  of the model(s)
 are obtained by continuation.
In this approach, the linearized problem is solved on a sequence of meshes until the
required accuracy is reached.
Also,  a redistribution of the mesh points is automatically performed to roughly
equidistribute the error. With this adaptive mesh selection procedure, the equations are
solved on a sequence of meshes until the successful stopping criterion is reached, where
the deviation of the collocation solution from the true solution is below a prescribed error
tolerance.

\section{Gauged Abelian complex scalar field models}
In this Section, all the models employed conform to the class of models \re{gen-action} with $\bt=0$. We have studied both nontopological and topological vortices of models featuring the usual
quadratic kinetic term, and since that one with symmetry-breaking potential is the $p=1$ Abelian Higgs (AH) model (see footnote $3$), we have described the corresponding model with no symmetry-breaking also
as a $p=1$ AH model. 
\subsection{$p=1$ models: General results }
Here, we have characterised the $U(1)$ gauged complex scalar model with the label $p=1$ because the Lagrangian of the model is formally that of the $p=1$ Abelian Higgs model, where the self-interaction potential
of the scalar field is not $a\ priori$ specified. Employing a symmetry-breaking Higgs potential, this is indeed the $p=1$ Abelian Higgs model, while using a potential that does not break the symmetry, the resulting model
supports non-topological vortices. (While in the Higgs case we have considered also the $p=2$ model, we have not considered the corresponding more nonlinear $p=2$ analogue with no symmetry-breaking.)

\subsubsection{The reduced Lagrangian and boundary conditions}

The simplest gauged spinning vortices are found in a model containing a single complex scalar field
$\varphi$
(or, equivalently a real
field doublet $\f^{\al}$ $(\al=1,2)$)
gauged with respect to a U(1) field $A_\mu$.
Its Lagrangian reads (here we follow the notation introduced in the previous Section)
\begin{eqnarray}
\label{n1}
{\cal L}^{(1)} = \la_1\,(D_{\mu}\f^{\al}D^{\mu}\f^{\al} )-V(|\f^{\al}|^2)-\frac{1}{4}\lambda_2 F_{\mu\nu}^2 
+\ka \vep^{\lambda \mu\nu}  A_\lambda F_{\mu\nu} \ ,
\end{eqnarray}
where 
$
$
 $V(|\f^{\al}|^2) $ is the scalar field potential, not yet specified as symmetry-breaking or otherwise,  
$\la_1$, $\la_2$, 
and  $\ka$ are coupling constants, which we keep unspecified for the sake of generality.

The field equations are found by taking the variation of (\ref{n1})
with respect to the gauge potential $A_\mu$ and the scalar field $\phi^\alpha$.
Of  particular  interest here are the equations for the U(1) field, 
\be
\label{Max1CS}
 \la_2\,\pa_{\mu}F^{\mu\nu}+2\ka \vep^{\nu\tau\la} F_{\tau\la} =-2\la_1\,(\vep\f)^{\al}D^{\nu}\f^{\al}\,,
\ee
the right hand side of which defines the electromagnetic current $j^{\nu}$.

The reduced one dimensional Lagrange density resulting from the imposition of symmetry presented in the previous section is 
%
%
\bea
\label{lagstt1}
r^{-1}L^{(1)} =\frac{1}{2}\la_2\left(\frac{a'^2}{r^2}-b'^2\right)
+\la_1\eta^2\left[\left(h'^2+\frac{a^2h^2}{r^2}\right)-b^2\,h^2\right]
+V(h^2)
+\frac{2\ka}{r} \left[(a-n)b'-ba'\right]~.
\eea
This equation features a single (real) scalar amplitude $h(r)$
and two U(1) gauge potentials, an electric $b(r)$ and a magnetic one $a(r)$.
%

The field equations result in three complicated ODEs
for the functions $a,b$ and $h$,
which  are solved subject to a set of boundary conditions 
compatible with finiteness of the energy and regularity 
of the solutions.
At the origin, one imposes
 \begin{eqnarray}
\label{cond1}
a(0)=n,~~b'(0)=0,~~h(0)=0. 
\end{eqnarray} 

The boundary conditions at infinity follow from the behaviour of the 
scalar field there.
For the version of the model exhibiting symmetry breaking,
 \begin{eqnarray}
\label{cond2}
h(r) \to 1,~~a(r) \to 0,~~{\rm and}~~ b(r) \to 0~~~~{\rm as}~~r\to \infty.
\end{eqnarray}
Otherwise, for non-topological vortices, 
 \begin{eqnarray}
\label{cond3}
h(r) \to 0,~~a(r) \to a_{\infty},~~{\rm and}~~ b(r) \to b_{\infty}~~~~{\rm as}~~r\to \infty~.
\end{eqnarray}
This difference in the boundary conditions for the gauge  potentials 
originates in the presence of the terms $a^2h^2$ and $b^2h^2$ in the corresponding energy functional (see Rel. (\ref{enp1})
below), which should vanish as $r\to \infty$. 
Also,  $a_{\infty}$ and $b_{\infty}$
are nonzero constants
(with $b_{\infty}$ identified with 
the frequency $\om$ of the scalar field in the gauge
decoupling limit).

It should be noted that for a nonzero electric potential, the presence of a Chern-Simons
term in the action is a prerequisite,
independently  of the asymptotics of the scalar field.
This can easily be seen by investigating the $\kappa=0$ limit of the
$b$-equation, written as
  \begin{eqnarray}
\label{eq-b1}
 \lambda_2 \left(r b b' \right)'=\lambda_2 r b'^2+2 \lambda_1 \eta^2 r b^2h^2.
\end{eqnarray}
Integrating (\ref{eq-b1}) from zero to infinity, it follows that a nonzero $b$ is not compatible with 
the requirements of finite energy.

 \subsubsection{Electric charge and angular momentum}
The electric charge is computed from  (\ref{defQ1h}),
where 
\be
\label{j01}
j^{\nu}=-2\la_1(\vep\f)^aD^{\nu}\f^a~.
\ee
Then one finds
\be
\label{defQ1hn}
Q_e =\int\,\left( \la_2\pa_{i}F_{i0}+2\ka\,\vep_{ij}F_{ij}\right)d^2x\,,
\ee
 which, when subjected to azimuthal symmetry, is
\bea
 Q_e&=&2\pi \int_0^\infty \left(\la_2\,r^{-1}(r\,b')'-4\ka\,\frac{a'}{r}\right)\,r dr\label{Q1azimf}
\\
& =&
2\pi \la_2\,[r\,b']_0^{\infty}-8\pi  \ka\,[a]_0^{\infty}= 8\pi  \ka (n-a_{\infty}) .
 \label{Q1azimfh}
\eea 
In deriving this result, we use the asymptotic
behaviour discussed above. 

To calculate the angular momentum, we consider the
$T_{i0}$ components of the stress-energy tensor of the model, 
as resulting from (\ref{metric-Tik})
\be
\label{Ti01}
 T_{i0}= \la_2\,F_{ij}F_{0j}+2\la_1\,D_i\f^a\,D_0\f^a~.
\ee
After imposition of azimuthal symmetry one finds
\be
\label{calJ1}
 {T^t}_\theta={\cal J}=-\la_2\,a'\,b'-2\la_1 \eta^2\, a\,b\,h^2 . 
\ee
To simplify this relation, one uses the Maxwell equation for the electric potential
\bea
\label{GLsym}
2\la_1 \eta^2  bh^2&=&\la_2\, \frac{(rb')'}{r}-4 \ka r^{-1} a' . 
\eea
Whence \re{calJ1} can be written as
\be
\label{calJ1f}
-{\cal J}=\la_2\left[a'b'+r^{-1}a\,(r\,b')'\right]-4\ka r^{-1} (a^2)' , 
\ee
leading to the following  expression of the total angular momentum
\bea
J= 2\pi \int_0^\infty{\cal J}\,rdr 
&=&-2\pi\la_2\int_0^\infty(r\,a\,b')'dr+4\pi \ka \int_0^\infty  (a^2)' dr 
\nonumber\\
&=&-2\pi \la_2[r\,a\,b']_0^{\infty}+4\pi \ka [ a^2]_0^{\infty}=
4\pi \kappa (a_{\infty}^2-n^2). 
\eea
This relation is evaluated subject to the asymptotic behaviour of the fields  defined above by \re{cond1}-\re{cond2}. 

For a model exhibiting symmetry breaking one finds the quadratic relation
\bea
\label{Jtop1}
 J
=
-4\pi \ka n^2= -\frac{1}{16 \pi \ka}Q_e^2,
\eea
while the
 relation for nontopological vortices is more complicated, with
\bea
\label{Jnontop1}
 J
= -nQ_e+\frac{1}{16\pi \ka}Q_e^2~.
\eea
Also, we notice that
 both the electric charge and angular momentum  are determined by the contribution of the
Maxwell-CS term only. Moreover, the presence of a standard Maxwell term in the Lagrangian is $not$
a prerequisite for the existence of solutions.

For completness, we give here the corresponding 
energy density functional of the gauged Abelian model:
\be
\label{enp1}
r^{-1}H^{} =
\frac{1}{2} \la_2\left(\frac{a'^2}{r^2}+b'^2\right)
+ \la_1\eta^2\left[\left(h'^2+\frac{a^2h^2}{r^2}\right)
+b^2\,h^2\right]
+V(h)~,
\ee
(note the absence of the Chern-Simons term in this relation).

\begin{figure}[h!]
\centering
\includegraphics[height=3in]{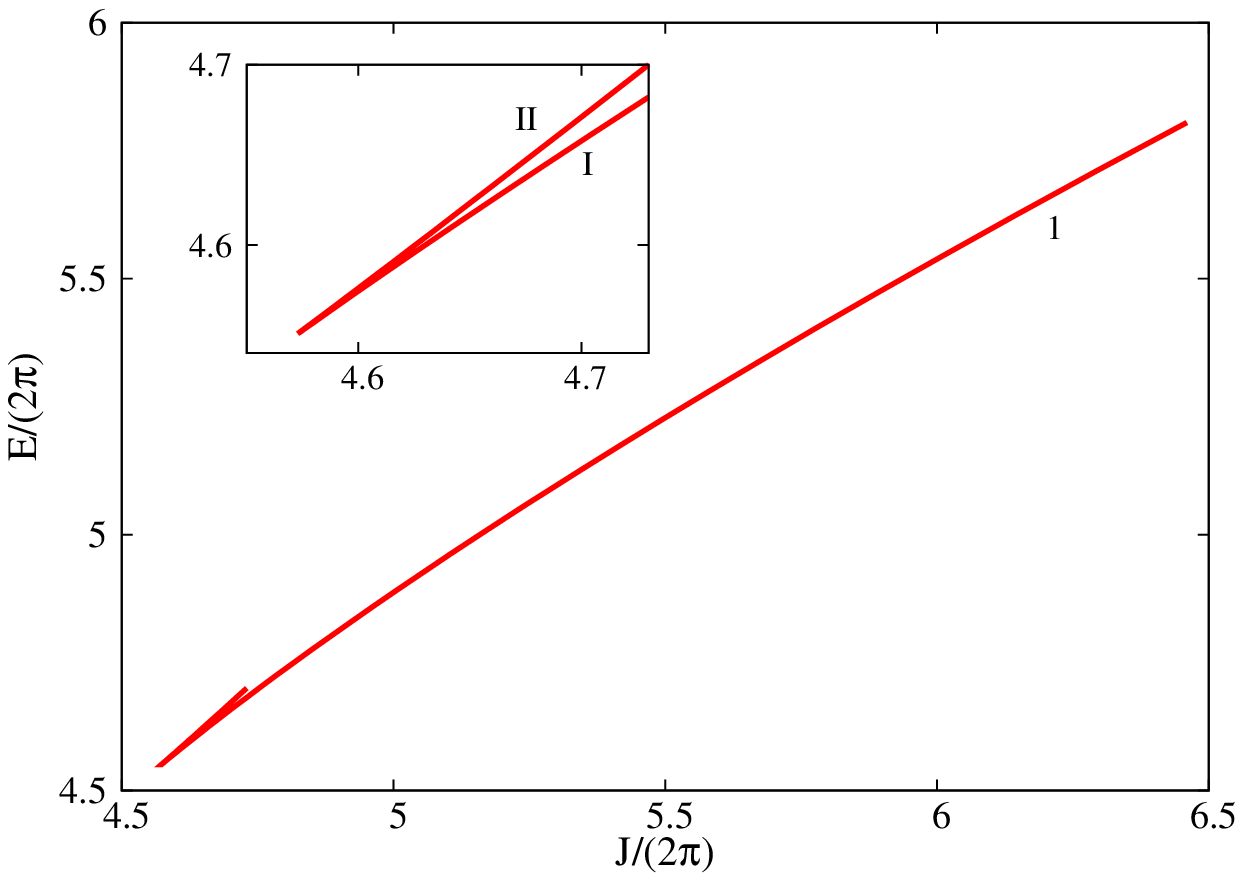}  
\includegraphics[height=3in]{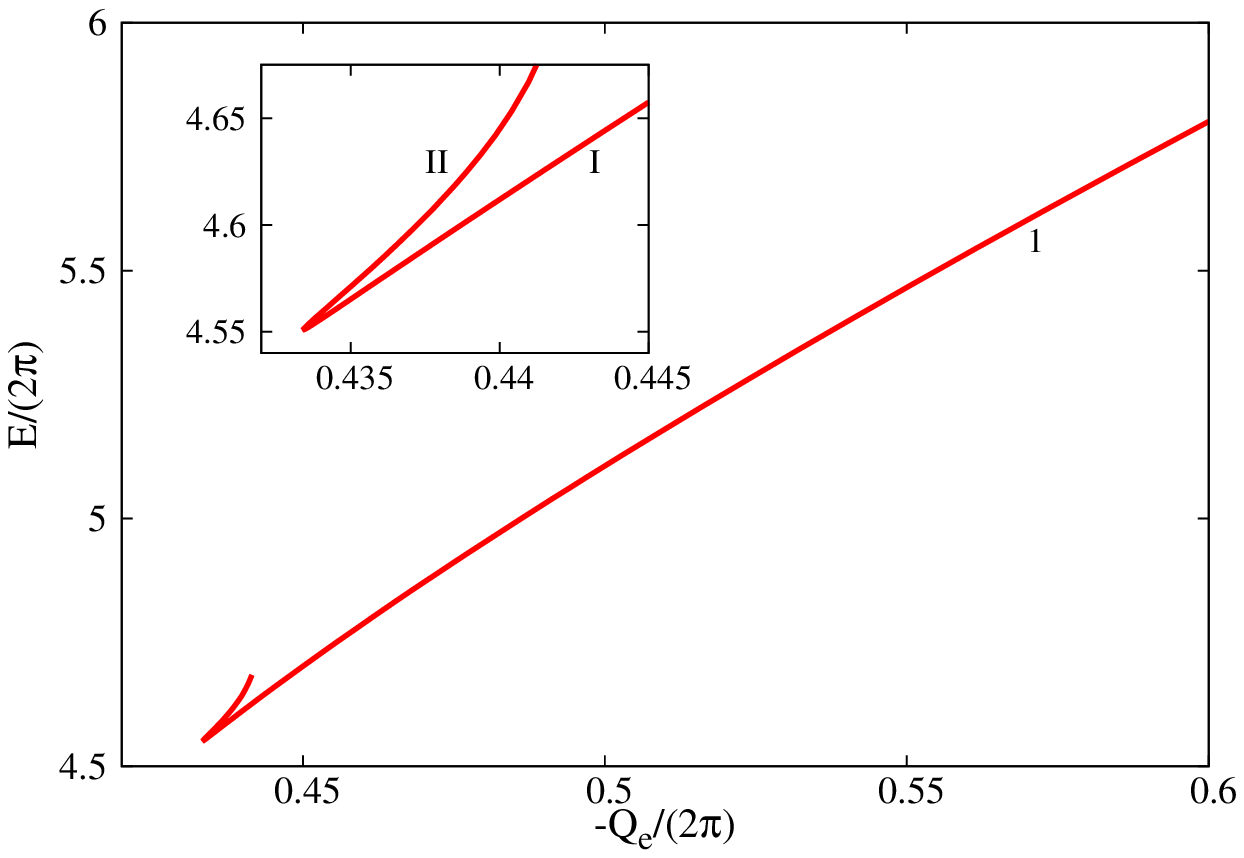}  

\caption{Energy $E$ is shown $vs.$ angular momentum $J$ 
and electric charge $Q_e$ 
for a family of $p=1$ gauged non-topological  vortices
with $c_1=2$, $c_2=-1$, $c_3=1.1$, $\lambda_1=1$, $\lambda_2=1/4$, and $\ka=0.1$. 
}
\label{fig1}
\end{figure} 
\subsubsection{ Non-topological vortices}

In contrast to the case of a Higgs (or a Goldstone) field in the next Subsection, 
the scalar field vanishes asymptotically, $h(r)\to 0$ 
(with $V \to 0$ in that limit),
such that this model does not possess any topological features.   
One remarks that this model possesses, however, a gauge-decoupling limit
which is found by suppressing the gauge field.
 This limit is discussed in the Appendix B.

In some sense, these are the $(2+1)$-dimensional counterparts of the gauged Q-balls
in four dimensions (see \cite{Radu:2008pp} for a review of their properties).
As in that case, 
one possible expression of the potential $V(|\f^{\al}|^2)$ 
which allows for spinning vortices with finite mass and angular momentum is
\begin{eqnarray}
\label{pot1}
 V(|\f^{a}|^2)=c_3 |\f^{\al}|^2+c_2|\f^{\al}|^4+c_1|\f^{\al}|^6
=c_3 \eta^2 h^2+c_2 \eta^4 h^4+c_1 \eta^6 h^6~,
\end{eqnarray}
with $c_i$ input parameters.
%

Some properties of these  $(2+1)$-dimensional non-topological 
gauged vortices were discussed in 
Ref. \cite{DeshaiesJacques:2006ae}.
The parameter space being very large, some properties of the solutions
appear to depend on the choice of parameters
$\lambda_i$,
$\kappa$ and 
$c_i$ of the theory.

After fixing these parameters,
the model still possesses two input constants, 
$a_{\infty}$ and $b_{\infty}$.
In our approach, the free parameter is taken to be $b_{\infty}$ -- the electric potential at infinity.
 It then follows that the corresponding value of $a_{\infty}$
is fixed by numerics, being unique for all solutions constructed so far
(although we cannot exclude the existence of excited configurations).
Note however, that $b_{\infty}$ cannot take arbitrary values and ranges over a finite
interval; the numerical integration becoming difficult at the limits
of that interval, with fast increasing values of $(E,a_{\infty})$.

Some results of the numerical integration are shown in Fig.~\ref{fig1}.
Those solutions have 
$c_1=2$, $c_2=-1$, $c_3=1.1$, $\lambda_1=1$, $\lambda_2=1/4$, and $\ka=0.1$.
As an interesting feature,
one notices that the minimal values for $Q_e$ and $J$ are approached 
for a critical configuration, which has also the minimal energy $E$.
Also, although the energy increases monotonically with $Q_e$,
one notices the existence of a degeneracy for a range of $Q_e$,
with the occurrence of a small secondary branch of solutions 
(labelled II
in Fig.\ref{fig1}).

\subsubsection{$p=1$ Abelian Higgs vortices: Topological}
These are the vortices of the $usual$ Abelian Higgs (AH) model typified by the potential
\begin{eqnarray}
 V(|\f^{a}|^2)= \la_0 (\eta^2-|\f^{a}|^2)^2
= \la_0\eta^4(1-h^2)^2\,.
\end{eqnarray}
in the Lagrangian \re{n1}.
In this case, the scalar field does not vanish as $r\to \infty$,
with the usual symmetry breaking scalar potential
and $h(r)\to 1$ asymptotically.

The one dimensional equations to be solved are those arising from the reduced Lagrangian \re{lagstt1}.
In contrast to the non-topological case typified by the potential \re{pot1} discussed above,
and the topological $p=2$ Abelian Higgs model \re{lagAH2} discussed below,
this model does not possess a gauge decoupling limit.


As expected, the presence of the CS term in \re{n1} results in electrically charged spinning vortices.
These vortex solutions are constructed numerically and their properties are investigated.

The boundary values of the solutions, following from the requirement of finite energy seen from \re{enp1}, are stated in (\ref{cond1}), (\ref{cond2}).
%

There is therefore no free parameter
characterising the solutions. The static energy density (\ref{enp1}) has no explicit
dependence on $\ka$, its integral nonetheless depends on $\ka$ through the dependence of the
functions $a(r)\ ,b(r)$ and $h(r)$ on it. Also the electric charge $Q_e(n)$ depends on $\ka$, so
one can plot $E(n)\ vs.\ Q_e(n)$ for fixed $n$ by varying $\ka$, $i.e.$ by varying the theory.
This is depicted in Fig~\re{fig2}, showing a monotonic increase of the energy with increasing electric charge.

\begin{figure}[h!]
\centering
\includegraphics[height=3in]{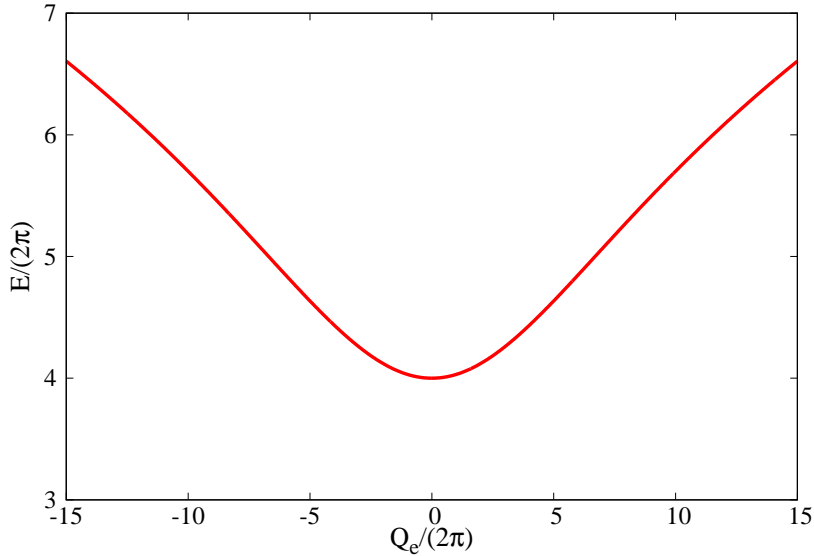}  
\caption{Energy $E$ vs. $Q_e$ for the $p=1$ AH system for $n=1$, $\lambda_2=\lambda_1=4.0$, $\lambda_0=2.0$, and $\eta=1.0$.} 
\label{fig2}
\end{figure} 

\subsection{$p=2$ Abelian-Higgs system }
\subsubsection{The reduced Lagrangian and boundary conditions}


Here, we shall consider a more general model consisting of the $p=2$ generalisation of the (usual) $p=1$ Abelian-Higgs model discussed above.
It is known that the qualitative properties of $D$-dimensional magnetic monopoles/vortices~\cite{Tchrakian:2010ar} (see footnote $3$)
of $p$-Yang-Mills--Higgs models (for allowed $(p,D)$ combinations) are similar. However, AH models characterised by different $p$ can display
differing quantitative features~\footnote{In \cite{Arthur:1998nh}, in the absence of a CS term and hence for purely magnetic vortices on $\R^2$,
it was found that the ``binding energies'' of
the vortices of the $p$-AH decrease with increasing $p$.}. This motivates our study here of the $p=2$ AH model, and a hybridised version of it,
in search of new properties when the CS term is present.
It turns out that in the presence of Chern-Simons
dynamics, some $p=2$ vortices exhibit new features similar to those observed above in the non-topological ($p=1$) model and the
Skyrme model below.




For technical reasons, related to the numerical convergence (stiff problem), a Maxwell term will be added to the pure $p=2$ Abelian Higgs system.
This term will subsequently be suppressed to leave the pure $p=2$ model under consideration, but some of the features revealed will
persist.

We start with reviewing the $p=2$ AH system.  
In the notation of \re{para}, the static Hamiltonian of the $p$-th Abelian Higgs models on $\R^2$ is (see Ref.~\cite{Tchrakian:2010ar} and references therein)
\bea
\label{pAHa}
{\cal H}^{(p)}_0&=&(1-|\f^{c}|^2)^{2(p-2)}\bigg(\la_2\left[(1-|\f^{a}|^2)\,F_{ij}+2\vep^{ab}\,(p-1)\,D_{i}\f^{a}D_{j}\f^{b}\right]^2+ \nonumber\\
 &&\qquad+4p(2p-1)\la_1(1-|\f^{c}|^2)^2\,|D_{i}\f^{a}|^2+2(2p-1)^2\la_0\,(1-|\f^{c}|^2)^4\bigg)~.
\eea
The $p=1$ model which results from (\ref{pAHa}) has been discussed above.

Of interest here is its $p=2$ generalization.
Using the shorthand notation
\bea
{\cal F}_{\mu\nu}&=&\left[(1-|\f^{\al}|^2)\,F_{\mu\nu}+2\,\vep^{\al\beta}D_{\mu}\f^{\al}D_{\nu}\f^{\beta}\right] , 
\label{cal2}
\nonumber
\\
{\cal F}_{\mu}^{\al}&=&(1-|\f^{\al}|^2)\,D_{\mu}\f^a , 
\label{cal1}
\\
{\cal F}&=&(1-|\f^{\al}|^2)^2 , 
\nonumber
\label{cal0}
\eea
the static Hamiltonian of the $p=2$ AH system can be written concisely as
\be
\label{ham0AH2}
{\cal H}^{(2)}_0=\la_2\,{\cal F}_{ij}^2+24\la_1\,|{\cal F}_{i}^a|^2+18\la_0\,{\cal F}^2,
\ee
 the corresponding Lagrangian being
\be
\label{lagAH2}
{\cal L}^{(2)}_0=-\la_2\,{\cal F}_{\mu\nu}^2+24\la_1\,|{\cal F}_{\mu}^a|^2-18\la_0\,{\cal F}^2 .
\ee
It is the static limit of \re{lagAH2} that will be used to derive the equations of motion for the ``putative'' electrically charged
solutions with $A_0\neq 0$.



Again, we augment \re{lagAH2} with the CS term which defines the (pure) $p=2$ CS-Higgs Lagrangian
\be
\label{lagAH2CS}
{\cal L}^{(2)}={\cal L}^{(2)}_0+\ka\,\vep^{\la\mu\nu}\,A_{\la}F_{\mu\nu}\, .
\ee

In practice, we will employ a hybridised version of \re{Max2CS} which is augmented by a Maxwell term with some coupling strength $\al$,
\be
\label{hybrid}
{\cal L}_{(\al)}^{(2)}=-\al\,F^2_{\mu\nu}+{\cal L}^{(2)}\,.
\ee
 This term is 
introduced mainly for purely technical reasons to simplify the numerical integrations.
After the solutions of the system \re{hybrid} are constructed, we take the limit $\al\to 0$ to yield the solutions to the $p=2$ system \re{lagAH2CS}.
That the solutions persist in the limit $\al\to 0$, $i.e.$ for the pure $p=2$ model is seen from Fig.~\re{fig_alpha_limit}.

\begin{figure}[h!]
\centering
\includegraphics[height=3in]{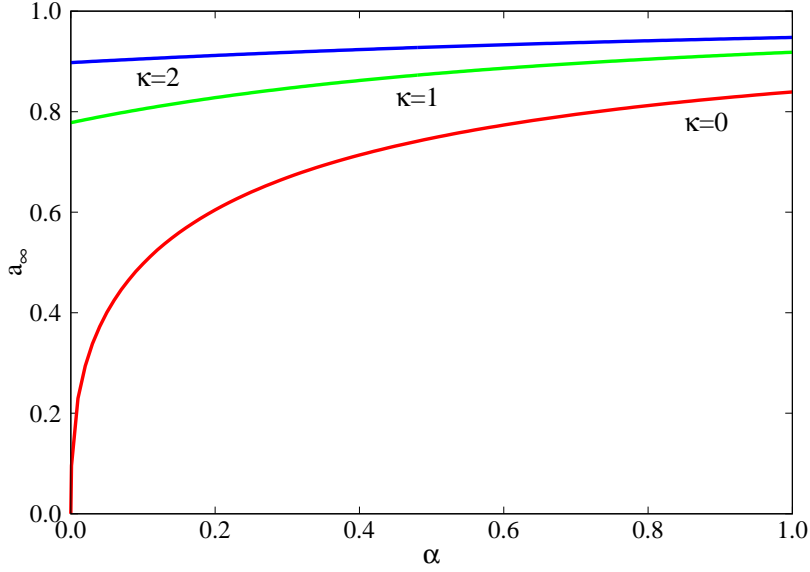} 
\caption{$a_\infty$ vs. $\alpha$ for the $F^2 + p=2$ AH system 
for $n=1$, $\lambda_2=\lambda_1=\lambda_0=1.0$, $b_\infty=0$, $\eta=1.0$ for several values of $\kappa$: $\kappa=0, 1, 2$.
}
\label{fig_alpha_limit}
\end{figure} 

The ensuing Maxwell equations are
\be
\label{Max2CS}
\al\,\pa_{\mu}F^{\mu\nu}+\la_2\,(1-|\f^{\al}|^2)\pa_{\mu}{\cal F}^{\mu\nu}+12\la_1\,(1-|\f^{\al}|^2)^2(\vep\f)^{\beta}D^{\nu}\f^{\beta}=-\frac12\ka\,\vep^{\nu\tau\la}F_{\tau\la} \ ,
\ee
which will be used later.

The reduced one dimensional Lagrangian resulting from imposition of symmetry is
\bea
\label{lagstt2}
r^{-1}L_{(\al)}^{(2)}&=&-2\al\,\left(\frac{a'^2}{r^2}-b'^2\right)-2\la_2\eta^4\left[r^{-2}\left([(1-h^2)a]'\right)^2-\left([(1-h^2)b]'\right)^2\right]\nonumber\\
&&\qquad\qquad\qquad\qquad-24\la_1\eta^6(1-h^2)^2\left[\left(h'^2+\frac{a^2h^2}{r^2}\right)-b^2\,h^2\right]\nonumber\\
&&\qquad\qquad\qquad\qquad-18\la_0\eta^8(1-h^2)^4+\frac{2\ka}{r}[(ab'-ba')-nb'] .
\eea
which we solve subject to the boundary values
\bea
&&\lim_{r\to 0}h(r)=0\ ,\ \quad\lim_{r\to 0}a(r)=n\ ,\ \quad\lim_{r\to 0}b'(r)=0 \ ,\label{p20}\\
&&\lim_{r\to\infty}h(r)=1\ ,\quad\lim_{r\to\infty}a(r)=a_\infty,\ ,\quad\lim_{r\to\infty}b(r)=b_\infty \ . \label{p2infty}
\eea
In contrast to the case of a $p=1$ AH model, here we see from  \re{p2infty} that
$a_\infty$ and $b_\infty$ are nonvanishing constants, in particular $b_\infty$ being a free parameter that allows us to vary the electric charge of the solutions within a concrete model (i.e., choice of the parameters in the Lagrangian). 
This can be gleaned by inspecting the static Hamiltonian corresponding to the Lagrangian \re{lagstt2}
\bea
\label{enp2}
r^{-1}H_{(\al)}^{(2)}&=&2\al\,\left(\frac{a'^2}{r^2}+b'^2\right)+2\la_2\eta^4\left[r^{-2}\left([(1-h^2)a]'\right)^2+\left([(1-h^2)b]'\right)^2\right]\nonumber\\
&&\qquad\qquad+24\la_1\eta^6(1-h^2)^2\left[\left(h'^2+\frac{a^2h^2}{r^2}\right)+b^2\,h^2\right]+ 18\la_0\eta^8(1-h^2)^4 \, ,
\eea
from which it is clear that finite energy does not require the constants $a_\infty$ and $b_\infty$ to vanish at infinity.
The solutions of the system \re{enp2} finite energy and topologically stable solutions (stabilised by the appropriate magnetic charge~\cite{Tchrakian:2010ar}) for with
\be
\label{cond4}
\lim_{r\to\infty}a(r)=a_\infty > 0\,.
\ee
Since the $F^2$ term does not introduce new fields, and since it is positive definite, it
suffices to consider the topological lower bound on the energy of the pure $p=2$ system alone.
The topological charges of $p$--Abelian-Higgs (AH) vortices
are the magnetic vortex numbers of the $p$-AH models introduced in Refs.~\cite{Burzlaff:1994tf,Arthur:1998nh}
and \cite{Tchrakian:2010ar}. The general expressions for these
topological charge densities are $total\ divergences$ on $\R^2$ which
subject to azimuthal symmetry take the simple expression~\cite{Arthur:1998nh}
\be
\label{axpvortex}
\vr^{(p)}=\frac{d}{dr}\left[(1-h^2)^{2p-1}\,a\right]\, ,
\ee
which is a $total\ derivative$ with respect to $r$. Hence the integral of \re{axpvortex} with the boundary values
\re{p20}-\re{p2infty} results in an $integer$ (vortex number) as required for topological stability.
Thus topological stability persists for the asymptotic value $a(\infty)>0$, \re{cond4}, for all $p$-AH models
with $p\ge 2$, but excluding $p=1$. Clearly, in the presence of the  Maxwell term $F^2$ the absolute minimum of the energy cannot be attained.

What is more important is that the solutions of the equations arising from the Lagrangian \re{lagstt2} feature the function $b(r)$ in addition to its first and second-order derivatives $b'$ and $b''$, as a result of which
$b_\infty$ is now a free parameter that characterises the solutions and cannot be fixed a priori. In the numerical process $b_\infty$ and $a_\infty$ are related, as shown in Fig.~\re{fig_alpha_limit_a_inf}. This enables the tracking of the energy $E$ with varying $Q_e$ and $J$. Note that when $\ka=0$, $b_\infty=0$ is fixed.

\begin{figure}[h!]
\centering
\includegraphics[height=3in]{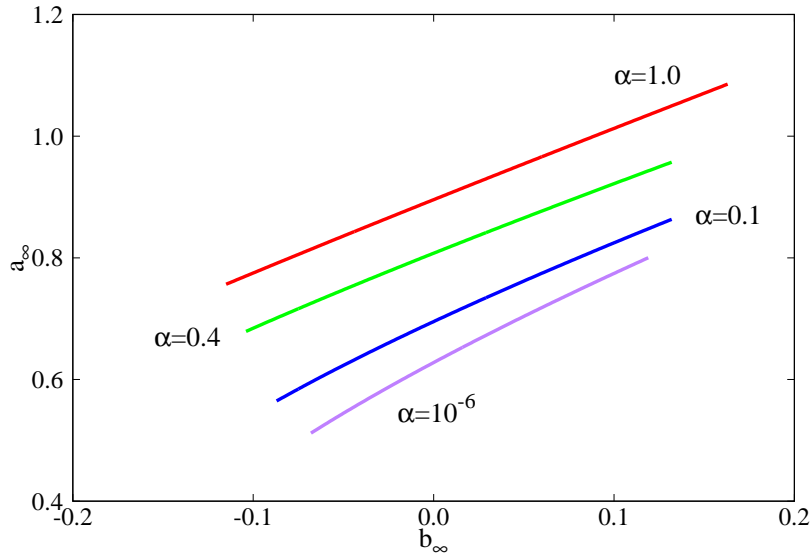}
\caption{$a_\infty$ vs. $b_\infty$ for the $F^2 + p=2$ AH system for $n=1$,  $\lambda_2=\lambda_1=\lambda_0=1.0$, $\ka=0.5$, and $\eta=1.0$ for several values of $\al$: $\al=1.0, 0.4, 0.1, 10^{-6}$.
}
\label{fig_alpha_limit_a_inf}
\end{figure} 

\subsubsection{Electric charge and angular momentum of the $p=2$ AH model}

Adopting the definition of the electric charge for the system \re{hybrid}
\be
\label{defQhyb}
\frac{1}{2}Q_e\stackrel{\rm def.}=\int\,\left(2\al\,\pa_iF_{i0}+2\la_2(1-|\f^{\al}|^2)\pa_{i}{\cal F}_{i0}-\ka\,\vep_{ij}F_{ij}\right)d^2x\,,
\ee
one has after symmetry imposition
\bea
\frac{1}{8\pi}Q_e&=&-\int\left(\al\,r^{-1}(r\,b')'+\la_2\,r^{-1}(r\,[(1-h^2)b]')'+\ka\,\frac{a'}{r}\right)\,rdr\label{Qazimhyb} \nonumber\\
&=&-\left[\al\,[r\,b']_0^{\infty}+\la_2\,[r\,[(1-h^2)b]']_0^{\infty}+\ka\,[a]_0^{\infty}\right]=\ka(n-a_\infty) . \label{Qazimfh}
\eea
Likewise, we have the $T_{i0}$ component of the hybrid model
\be
\label{Ti0hyb}
T_{i0}=4\al\,F_{ij}{F_{0}}^j+4\la_2\,{\cal F}_{ij}{{\cal F}_{0}}^j+48\la_1\,{\cal F}_{i}^a{\cal F}_{0}^a~,
\ee
which after imposition of symmetry yields
\be
\label{calJhyp}
\frac{1}{4}\,{\cal J}=\al a'b' +\la_2 \eta^4\,[(1-h^2)a]'[(1-h^2)b]'+12\la_1 \eta^6\,(1-h^2)^2a\,b\,h^2~.
\ee
leading to the final expression of the angular momentum
\bea
\frac{1}{8\pi}\,J&=&\al[r\,a\,b']_0^{\infty}+\la_2 \eta^4[r[(1-h^2)a]'[(1-h^2)b]']_0^{\infty}+\frac12\ka\,[a^2]_0^{\infty}=\frac12\ka(a_\infty^2-n^2)\, .\label{Jhyp}
\eea

\subsubsection{The solutions}

In Fig.~(\ref{fig1n}) we represent the energy $E$ of the solutions as a funtion of the electric charge $Q_e$ for a certain system with nonvanishing CS coupling constant ($\ka=0.5$). The free parameter here is the value of $b$ at infinity, which is no longer a gauge freedom. We clearly observe that minimal energy occurs at a nonvanishing value of the electric charge, so the uncharged solution is not energetically favoured in this model.

\begin{figure}[h!]
\centering
\includegraphics[height=3in]{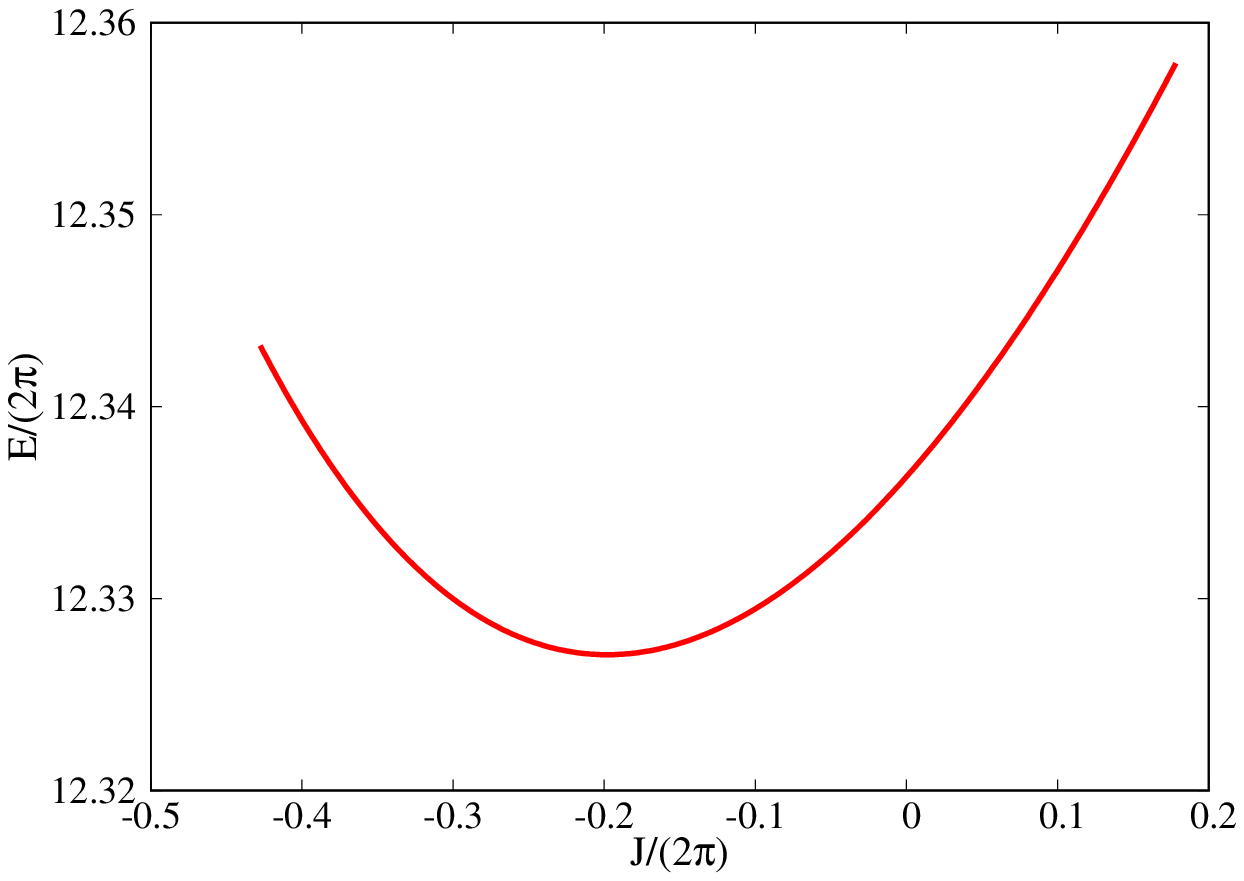}
\includegraphics[height=3in]{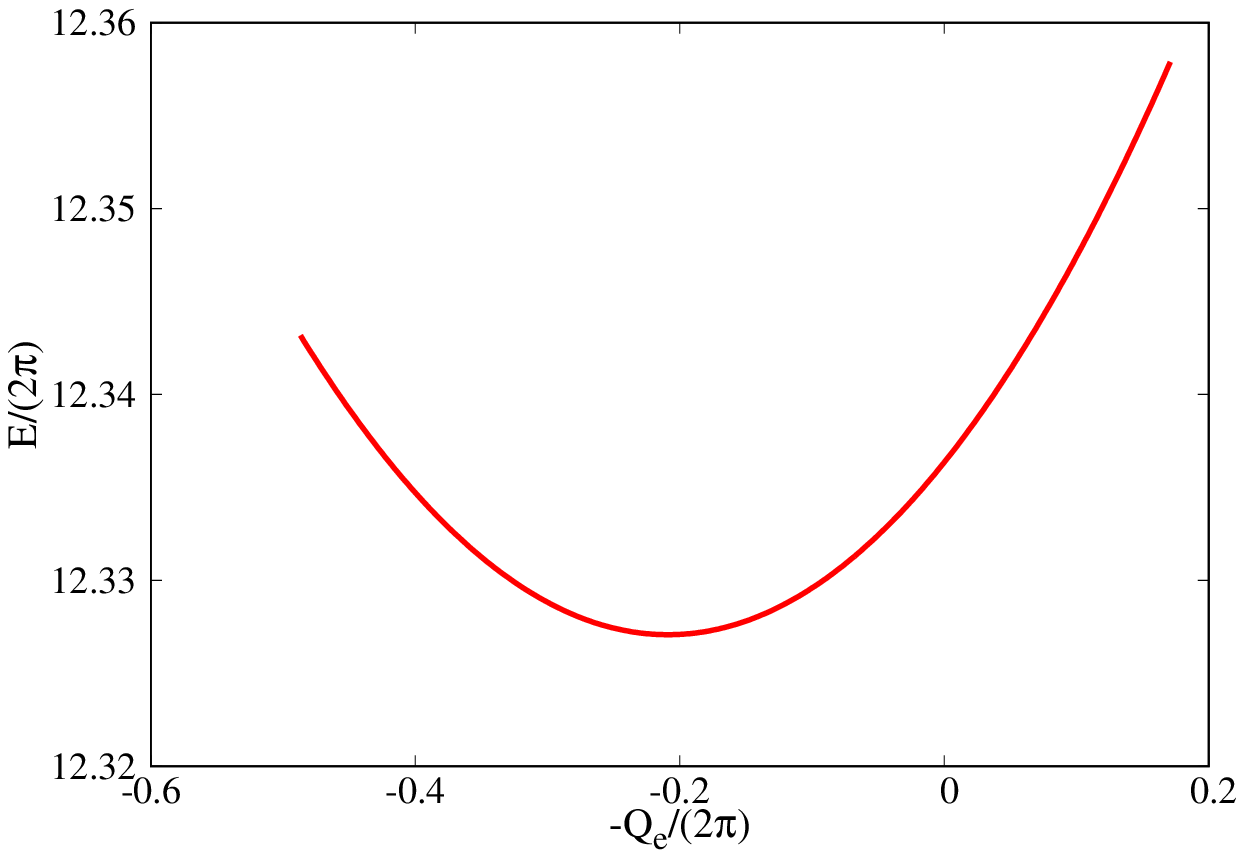}
\caption{Energy $E$ $vs.$
angular momentum $J$ and electric charge $Q_e$ for the $F^2 + p=2$ AH system for $n=1$, $\alpha=1.0$, $\lambda_2=\lambda_1=\lambda_0=1.0$, $\ka=0.5$, and $\eta=1.0$.
}
\label{fig1n}
\end{figure} 

Related to this fact, the relation between the energy $E$ and the absolute value of the angular momentum $|J|$ ceases to be monotonically increasing, the minimal energy being reached at a nonvanising angular momentum, as seen in Fig.~(\ref{fig1n}). Moreover, given a concrete theory, both the electric charge and the angular momentum are bounded quantities supported on finite ranges of $Q_e$ and $J$.

\section{Abelian Skyrme model in $2+1$ dimensions: Topological}
Unlike in the case of the Abelian gauged complex scalar models considered in Section {\bf 3}, where both topological and non-topological vortices were studied, here, in the case of the Abelian gauged $O(3)$ sigma (Skyrme)
models we are exclusively concerned with topological solitons.
The models employed in this Section conform to the class of models \re{gen-action} with $\gamma=0$.
\subsection{The model}
The Lagrangian of the simplest Abelian Skyrme model~\cite{Schroers:1995he} in $2+1$ dimensions is
\be
\label{MaxO3}
{\cal L}_0=-\frac14\,\la_2\,F_{\mu\nu}F^{\mu\nu}+\frac12\,\la_1\,|D_{\mu}\ta^a|^2-\frac12\la_0\,V[\ta^3]\, ,
\ee
with the spacetime index $\mu=(0,i)$, $i=1,2$.
In the static limit, \re{MaxO3} supports topologically stable~\footnote{It turns out that for $\la_0=1$ and $V=(\f^3-1)^2$, these vortices saturate the topological (magnetic) lower bound.} magnetic vortices first constructed in Ref.~\cite{Schroers:1995he}.

The most general Abelian Skyrme model is
\be
\label{MaxSk}
{\cal L}_1=-\frac14\,\la_2\,F_{\mu\nu}F^{\mu\nu}-\frac18\,\la_3\,|D_{[\mu}\ta^aD_{\nu]}\ta^b|^2+\frac12\,\la_1\,|D_{\mu}\ta^a|^2-\frac12\la_0\,V[\ta^3] \, ,
\ee
having added the quartic kinetic term $|D_{[\mu}\ta^aD_{\nu]}\ta^b|^2$ in which the index notation $[\mu\,\nu]$ implies antisymmetrisation in $\mu$ and  $\nu$. To \re{MaxSk} we add the Abelian Chern-Simons (CS) term \re{lcs},
and we study the system
\be
\label{Lagqqcs}
{\cal L}={\cal L}_1+{\cal L}_{\rm CS} , 
\ee
which supports both electric charge and angular momentum.

We will choose $V[\ta^3]$ to be the usual ``pion mass'' potential
\be
\label{pion}
V[\ta^3]=(1-\ta^3)\,.
\ee
The Maxwell equation resulting from the Lagrangian \re{Lagqqcs} is
\be
\label{maxqqudquar}
\la_2\pa_{\mu}F^{\mu\nu}-\la_3D^{[\mu}\ta^{\al}D^{\nu]}\ta^{\bt}\ (\vep\ta)^{\al}D_{\mu}\ta^{\bt}+\la_1(\vep\ta)^{\al}D^{\nu}\ta^{\al}=-2\ka\,\vep^{\nu\rho\si}F_{\rho\si} .
\ee

Exploiting \re{di0fal}-\re{di0f4} leads to the reduced one dimensional Lagrangian density
\bea
\label{redLag}
r^{-1}\,L&=&\frac12 \, \la_2\,\left(\frac{a'^2}{r^2}-b'^2\right)
+\la_3\left(\frac{a^2}{r^2}
-b^2\right)\sin^2 f f'^2 
\nonumber
\\
&&+\frac12\,\la_1\,\left[\left(\frac{a^2}{r^2}-b^2\right)\sin^2f+f'^2 \right]+
\frac12\,\la_0(1-\cos f)+\frac{2\ka}{r}[(ab'-ba')-nb'] ,
\eea
which we solve subject to the boundary values
\bea
&&\lim_{r\to 0}f(r)=\pi\ ,\ \quad\lim_{r\to 0}a(r)=n\ ,\ \quad\lim_{r\to 0}b'(r)=0 \ ,\label{sk0}\\
&&\lim_{r\to\infty}f(r)=0\ ,\quad\lim_{r\to\infty}a(r)=a_\infty,\ ,\quad\lim_{r\to\infty}b(r)=b_\infty \ , \label{skinfty}
\eea
where $a_\infty$ is not necessarily zero and $b_\infty$ is a free parameter that allows us to vary the electric charge of the solutions within a concrete model (i.e., choice of the parameters in the Lagrangian).
Notice that $a_\infty$ is numerically related to $b_\infty$.

\subsection{Electric charge and angular momentum}
The electric current, in terms of the scalar matter fields, is
\be
\label{jnu}
j^{\nu}=\la_3D^{[\mu}\ta^{\al}D^{\nu]}\ta^{\bt}\ (\vep\ta)^{\al}D_{\mu}\ta^{\bt}-\la_1(\vep\ta)^{\al}D^{\nu}\ta^{\al} .
\ee 
In terms of the static azimuthally symmetric fields \re{axAi0}, the electric charge is
\bea
\label{Q1azim}
Q_e=\int j_0\,d^2x&=&-2\pi\int\left(\la_2\,r^{-1}(r\,b')'+4\,\ka\,\frac{a'}{r}\right)\,rdr\nonumber\\
&=&-2\pi\left[\la_2\,\left[(r\,b')\right]_{0}^{\infty}+4\,\ka\,\left[a\right]_{0}^{\infty}\right]=8\pi\,\ka(n-a_\infty)\, .
\eea

To calculate the angular momentum of the model described by the Lagrangian \re{MaxSk} we need the relevant component of the stress tensor
\be
\label{ti0sk}
T_{i0}=\la_2\,F_{ij}{F_{0}}^j+\frac12 \la_3\,(D_{[i}\ta^aD_{j]}\ta^b)(D_{[j}\ta^aD_{0]}\ta^b)-\la_1\,D_{i}\ta^aD_{0}\ta^a ,
\ee
which when subjected to azimuthal symmetry reduces to 
\be
\label{ti0sksym}
T_{i0}=r^{-1}\left\{\la_2\,a'b'+ab\,P^2\left[\la_1+2\la_3(P'^2+Q'^2)\right]\right\} ,
\ee
leading to the angular momentum density
\be
\label{jz}
{\cal J}=\left\{\la_2\,a'b'+ab\,P^2\left[\la_1+2\la_3(P'^2+Q'^2)\right]\right\} .
\ee

Now the Gauss Law equation arising from the variation of the Lagrangian \re{MaxSk}, when subjected to this symmetry is
\be
\label{gauss}
b\,P^2\left[\la_1+2\la_3(P'^2+Q'^2)\right]=r^{-1}[\la_2(rb')'+4\,\ka\,a'] ,
\ee
whence \re{jz} simplifies to
\be
\label{calJ1fs}
\,{\cal J}=r^{-1}\left[\la_2(rab')'+2\ka(a^2)'\right] ,
\ee
yielding the final expression of the angular momentum
\bea
J=2\pi\int{\cal J}\,rdr
&=&2\pi\left[\la_2[r\,a\,b']_0^{\infty}+2\ka\,[a^2]_0^{\infty}\right]\nonumber\\
&=&4\pi\ka[a_\infty^2-n^2]\label{J1s}\,.
\eea

It is clear from \re{J1s} that the angular momentum vanishes in the absence of the CS term, $i.e.$ when $\ka=0$. This
is known from the work of Ref.~\cite{Gladikowski:1995sc}, namely that static $U(1)$ gauged Skyrmions  do not have angular momentum, inspite
of the fact that in the gauge decoupling limit they do support $J$ as known from the work of Ref.~\cite{Piette:1994mh}. We will return to the last example in Appendix {A.3}. The question
of angular momentum of the (gauged and ungauged) Skyrmions of the $O(3)$ Skyrme model in 2+1 dimensions markedly contrasts with that of the $O(4)$ Skyrme model in 3+1 dimensions. There,
the $U(1)$ gauged Skyrmion of the $O(4)$ sigma model in $3+1$ dimensions does spin as shown in \cite{Radu:2005jp}, as also does the (ungauged) Skyrmion~\cite{Battye:2005nx}.

\subsection{The solutions}
For these solutions $b_\infty$, the asymptotic value of $b(r)$, turns out to be a free parameter.
Through the numerical process $a_\infty$, the asymptotic value of $a(r)$, is related to $b_\infty$. In Fig.~(\ref{fig45}) the numerical relation between $a_\infty$ and $b_\infty$ is shown for several values of $\kappa$.
Now the value of the energy $E$ depends both on $a_\infty$ and $b_\infty$, while $Q_e$  and $J$ explicitly
depend on $a_\infty$. Thus the dependence of $E$ on $Q_e$ and on $J$ can be tracked by varying $b_\infty$. These are depicted in Fig.~\re{fig2n} and  Fig.~\re{fig2p} respectively.

\begin{figure}[h!]
\centering
\includegraphics[height=3in]{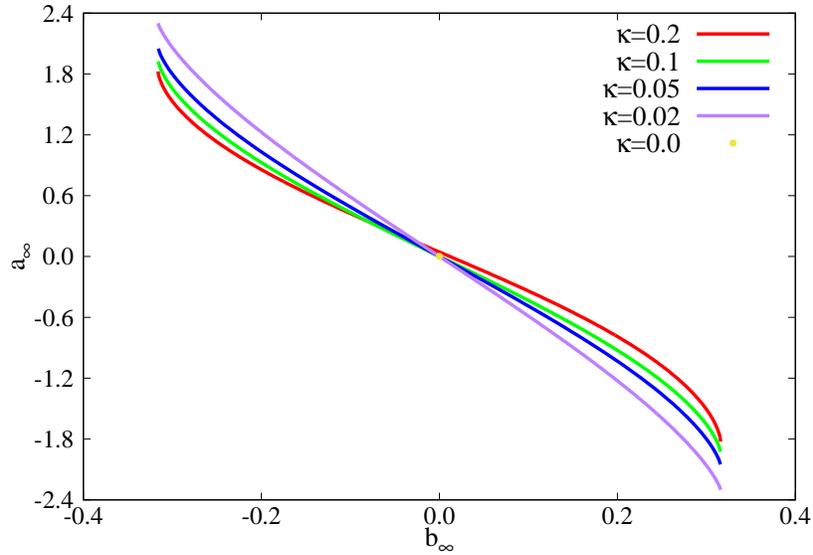}  
\caption{Asymptotic value $a_\infty$ vs. asymptotic value $b_\infty$ for the Abelian Skyrme model for $n=1$, $\lambda_0=0.2$, $\lambda_1=1.0$, $\lambda_2=0.01$, $\lambda_3=0.5$ and several values of $\ka$.  
}
\label{fig45}
\end{figure} 

The situation for this model is similar to that of the one in Section {\bf 3.2}. The fact that $b$ at infinity constitutes a free parameter of the theory allows charged configurations with less energy than the corresponding uncharged one. This fact is shown in Fig.~(\ref{fig2n}). There we can see that the effect exists for any nonvanishing value of $\ka$. When the energy is represented as a funtion of the angular momentum, Fig.~(\ref{fig2p}), we observe that for a given value of the angular momentum there are two different solutions with different mass (and different electric charge). Again both the charge and the angular momentum are bounded, as seen in Figs.~(\ref{fig2n}) and (\ref{fig2p}).

\begin{figure}[h!]
\centering
\includegraphics[height=3in]{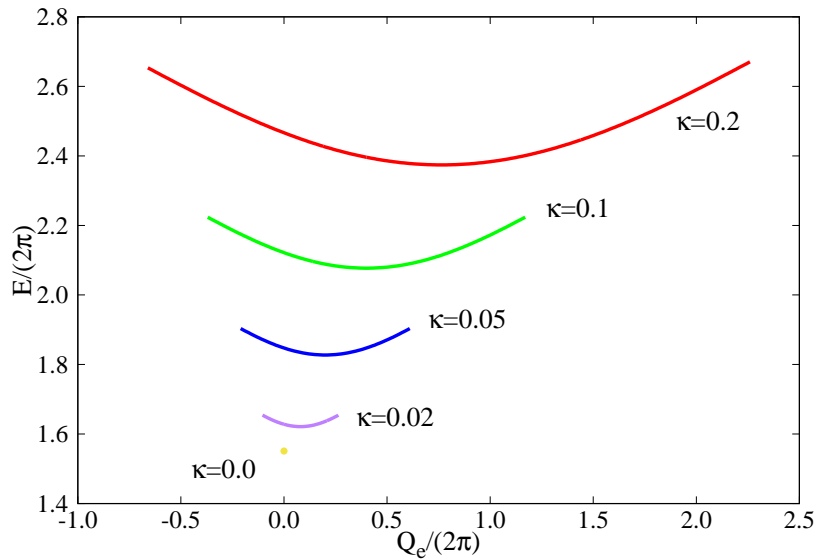}
\caption{Energy $E$ vs. electric charge $Q_e$ for the Abelian Skyrme model for $n=1$, $\lambda_0=0.2$, $\lambda_1=1.0$, $\lambda_2=0.01$, $\lambda_3=0.5$ and several values of $\ka$.  
}
\label{fig2n}
\end{figure} 

\begin{figure}[h!]
\centering
\includegraphics[height=3in]{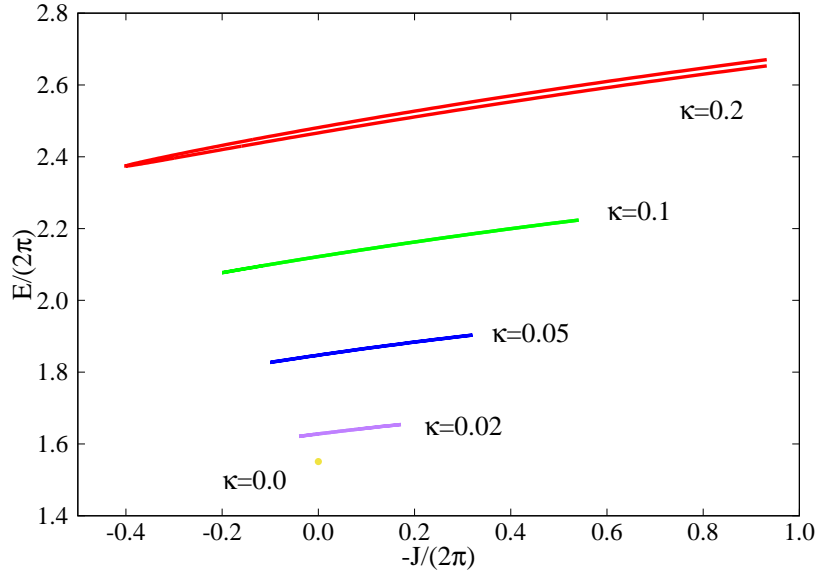}  
\caption{Energy $E$ vs. angular momentum $J$ for the Abelian Skyrme model for $n=1$, $\lambda_0=0.2$, $\lambda_1=1.0$, $\lambda_2=0.01$, $\lambda_3=0.5$ and several values of $\ka$.
}
\label{fig2p}
\end{figure} 

\section{Summary and discussion}

The overriding aim of this work was to investigate the effect Chern-Simons (CS) dynamics has on
various classes of solitons
in $2+1$ dimensions.
 The focus of our interest was on the effect that CS dynamics has on
the dependence of the energy $E$ on the electric charge $Q_e$ 
and the angular momentum $J$ of the vortices of $U(1)$ gauged scalar field theories.

In the present work we have studied models of $U(1)$ gauged complex scalar fields and $O(3)$ Skyrme fields in $2+1$ dimensions, in the presence of the CS term.
The vortices of these models support $Q_e$ and $J$ only when the CS term is present.
In the case of models featuring symmetry-breaking dynamics, namely the Abelian-Higgs (AH) models and the $O(3)$ Skyrme model, the vortices (solitons) are topologically stable
(prior to the introduction of the CS term), while the (non-topological) vortices of models with no symmetry-breaking, they are gauged $Q_e$-vortices whose stability has its source in its dependence on the angular velocity/momentum.

While it was long known that CS dynamics endows the gauged topological vortices with electric charge $Q_e$ and angular momentum $J$, the dependence
of the mass-energy of these two global quantities was not investigated. This has been done here, and it is shown that the dependence
of the energy on $Q_e$ and $J$ is not monotonic. This contrasts with the electrically charged solitons of the familiar $SO(3)$ gauged Higgs solitons in $3+1$ dimensions,
namely the Julia-Zee dyons, where the energy increases monotonically with $Q_e$.

The main result of this work is that of the non-standard dependence of the energy $E$ on the electric charge $Q_e$ and the angular momentum $J$. Most remarkably, $E$ can decrease with increasing $Q_e$ in some regions of the parameter
space, in contrast to the $usual$ monotonic increase of $E$ with $Q_e$. Also, the dependence of $E$ on $J$ turns out to be non-standard, contrasting with the $usual$ linear relationship $J\propto Q_e$.

These new features are observed in models where the solutions allow for nonzero asymptotic values of the magnetic function $a(r)$, $i.e.$ $a(\infty)=a_{\infty}\neq0$. In turn, such solutions are a result of the occurrence of the
nonzero asymptotic values of the electric function $b(r)$, $i.e.$ $b(\infty)=b_{\infty} \neq 0$, which results from the equations of motion arising from the Lagrangian. 
Most importantly, the constant $b_{\infty}$ turns out to be a free parameter
characterising the solutions, and since $E$ depends both on $a_{\infty}$ and $b_{\infty}$, and $Q_e$ and $J$ explicitly depend on $a_{\infty}$, the dependence of $E$ on $Q_e$ and $J$ can be tracked. 
This situation obtains for 
$i)$
models supporting non-topological vortices,
$ii)$
the $p=2$ AH
model (see Fig.~\re{fig1n}  and
$iii)$
 the gauged Skyrme model (see Figs.~\re{fig2n} and \re{fig2p}).
However, it is absent in the case of the $p=1$ (usual) AH model.   

In Appendices {\bf B.1,\ B.2,\ B.3}, we have considered the gauge decoupling limits of these models. It turns out that in this limit $J$ is supported by the non-topological vortices of the complex scalar models and the topological
vortices of the $O(3)$ Skytrme nmodel. In the case of the topological vortices of complex scalar (AH) models, our numerical results for the $p=2$ AH model are not conclusive but $J$ is likely not supported.

Most of the configurations in this work possess generalizations in four spacetime dimensions\footnote{
A review of spinning solitons in various $d=3+1$ models can be found in \cite{Radu:2008pp}.}.
Thus it is interesting to contrast these two cases.
The first observation is that no Maxwell Chern-Simons terms exists in $d=3+1$ spacetime dimensions.
However, in that case, the electric field decays sufficiently slowly so as to allow for finite mass spinning solutions
with a Maxwell term only in some models with gauged scalar fields.

Starting with the spinning $U(1)$ gauged vortices with a non-topological scalar field, we notice the existence  
of $d=3+1$ counterparts with many similar properties  \cite{Radu:2008pp}.
However, the total angular momentum of those charged Q-ball solutions is proportional to the electric charge,
$J=n Q_e$, with $n$ an integer, the winding number.
The picture is very different for a Higgs-like complex scalar field exhibiting symmetry-breaking dynamics, 
in which case we are not aware of any finite mass, 
spinning particle-like solution in $d=3+1$ dimensions. 
Also, note that there exist no $U(1)$ gauged Higgs--Chern-Simons densities~\cite{Tchrakian:2015pka} in even dimensional spacetimes, which would have been necessary
to enable spin in this case.
The picture is different for models with non-Abelian gauge fields, 
but that is outside of the scope of this work.

The spinning solutions of the Abelian Skyrme model in Section {\bf 3} also possess generalisations in $3+1$ dimensions, 
for the $U(1)$ gauged $O(4)$ Skyrme model. 
An important difference being that in that case there is no Chern-Simons term in the Lagrangian,
but nonetheless the axially symmetric solutions still support angular momentum. 
The corresponding properties of the solutions are discussed in \cite{Radu:2005jp}, where it was seen that the angular momentum is related linearly to the electric charge~\footnote{This contrasts with the 2+1 dimensional case at hand where the angular momentum
is generically proportional to the $square$ of the electric charge. One should not seek too deep a reason for this difference between $2+1$ and $3+1$ dimensions, since the definitions of the electric charge in each case is different, in the former case
being reliant on the presence of the Chern-Simons term, and the latter on electric-magnetic duality.}.   Also in that model, the energy increases monotonically with increasing
electric charge as shown in \cite{Piette:1997ny}, which is probably due to the absence of Chern-Simons dynamics. 
 Nonzero angular momentum persists in
the gauge decoupling limit of this model, namely the $usual$ Skyrme model, 
as shown in \cite{Battye:2005nx} and briefly recovered in Appendix {\bf B.3} below.

Perhaps the most important conclusion from the results in this paper pertaining to the $U(1)$ gauged scalar field model with Maxwell--Chern-Simons dynamics, is the analogy with the results pertaing to non-Abelian Higgs models with Yang-Mills--Chern-Simons dynamics
presented in Ref.~\cite{Navarro-Lerida:2014rwa}. There, we have studied $SO(5)$ and SU(3) gauged Higgs models featuring ``Chern-Simons'' dynamics, where the ``Chern-Simons'' densities employed are what we have referred to as
Higgs--Chern-Simons~\cite{Tchrakian:2010ar,Radu:2011zy,Tchrakian:2015pka} densities. (As it happens the magnetic (topological) charge in the $SO(5)$ model was zero while in the $SU(3)$ model it was non-zero, in both cases with non-zero $Q_e$.)
It was observed there, like here, that the minimum of the energy did not always coincide with the $Q_e=0$ configuration. We maintain that this effect is due to the presence of Chern-Simons dynamics in both cases, which is remarkable since the Higgs--Chern-Simons (HCS)
densities in even dimensions are $gauge\ invariant$ in contrast to the Chern-Simons and Higgs--Chern-Simons densities in odd dimensions, which are $gauge\ variant$. Inspite of the different gauge transformation properties of the HCS densities in even and odd dimensions,
we observe the same dynamical effect in both cases. (It should be added that this analogy is not complete in relation to the dependence of the energy on the angular momentum. 
This is due to the well known absence of a global angular momentum for non-Abelian Higgs
configurations with a net magnetic charge
\cite{Navarro-Lerida:2014bja},
\cite{VanderBij:2001nm}.

 \vspace*{0.5cm}
\noindent{\textbf{~~~Acknowledgements.--~}} 
D.H.T. would like to thank for hospitality at the
Albert-Einstein-Institut (AEI), Golm, where some of this work was
carried out.
 E.R. acknowledges funding from the FCT-IF programme.
F. N.-L. acknowledges funding from Complutense University under project PR26/16-20312.
This work was also partially supported 
by  the  H2020-MSCA-RISE-2015 Grant No.  StronGrHEP-690904, 
and by the CIDMA project UID/MAT/04106/2013.


\appendix 
\setcounter{equation}{0} 
\renewcommand{\theequation}{A.\arabic{equation}}
 \section{ Conserved charges}


Given a model with  Lagrangian density ${\cal L}$,
the energy-momentum tensor $T_{\mu \nu}$ of the solutions
is most easily 
defined by introducing the spacetime metric $g_{\mu\nu}$ into
the action and assuming it to be arbitrary (see $e.g.$ Ref. \cite{Rubakov:2002fi}).
Then $T_{\mu \nu}$  (which is directly symmetric and gauge invariant)  is obtained
by differentiating the density of the action with respect to the metric\footnote{Note that, as usual, the CS terms in (\ref{gen-action})
do not contributes to the
 energy-momentum tensor.}:
\begin{eqnarray}
\label{metric-Tik}
 T_{\mu\nu}
=  \frac{2}{\sqrt{-g}}
\frac{\delta (\sqrt{-g} {\cal L)}}{\delta g^{\mu\nu}}.
\end{eqnarray}
(note that the metric $g^{\mu\nu}$ is set equal to the Minkowski metric $after$
differentiation).
For configurations whose energy-momentum tensor does not depend on both $\theta$ and $t$,
 $T_{t}^t$ and $T_\theta^t \equiv {\cal J}_z$ corresponds to 
the energy density and angular momentum density, respectively.
The total mass-energy $E$ and total angular momentum $J$ are the integral of these quantities,
\begin{eqnarray}
\label{EJ}
E=2\pi \int_0^\infty dr~r T_t^t,~~J=2\pi \int_0^\infty dr~ r T_\theta^t  .
\end{eqnarray}

The solutions possess also an electric charge
whose defintion is based on the Maxwell equation
\be
\label{EL1}
 \la_2\frac{1}{\sqrt{-g}}\pa_{\mu}(\sqrt{-g}F^{\mu\nu} )+2\ka \vep^{\nu \rho\sigma}F_{\rho \sigma} =j^{\nu}
\ee
where
\begin{eqnarray}
j^{\mu}=\frac{1}{\sqrt{-g}}
\frac{\delta (\sqrt{-g} {\cal L}_{S})}{\delta A_{\mu }} \, ,
\end{eqnarray} 
where ${\cal L}_{S}$ is the part of the Lagrangian density different from the Maxwell and the CS terms. 

The electric charge is the volume integral of $j^0=j_0=\rho$
\be
\label{defQ1h}
Q_e\stackrel{\rm def.}= 
\int j_0\,d^2x=2\pi \int_0^\infty dr~r j_0.
\ee

\section{The gauge decoupling limits: Spinning vortices}
\setcounter{equation}{0} 
\renewcommand{\theequation}{B.\arabic{equation}}
In these Appendices, we study the gauge decoupling limits of the pure $p=1$ (non-topological), pure $p=2$ Higgs, and the $U(1)$ gauged Skyrme models, in the absence of
the Chern-Simons term since the latter vanishes in this limit. Our motive here is to consider the spin of these vortices in the stationary limit.

The Lagrangians in the gauge decoupling limits follow from the replacements
\be
\label{gaugedec}
a(r)\to n\ ,\qquad b(r)\to\om ,
\ee
and in this limit the axially symmetric Ans\"atze \re{h} and \re{fa} for the scalar fields are upgraded by replacing~\footnote{In the gauged models, the angular momentum is calculated in terms of the static fields due to the presence of the Abelian field.}
the unit vector $n^{\al}$ in them with
\begin{eqnarray}
\label{nhat}
\hat n^{a}=\left[
\begin{array}{c}
\cos (n\ta-\om t)\\
\sin  (n\ta-\om t)   
\end{array}
\right] , 
\end{eqnarray}

 \subsection{Spinning Q-vortices}

The simplest model possessing spinning solitons in $2+1$ dimensions contains a single complex scalar field
$\varphi$
(or, equivalently a
scalar doublet $\f^{a}$ $(a=1,2)$).
Its Lagrangian reads
(here we follow the notation of \re{para})
\begin{eqnarray}
\label{n1n}
{\cal L} =  \la_1\,(\partial_{\mu}\varphi^*\partial^{\mu}\varphi )-V(|\varphi|^2) 
  =  \la_1\,(\partial_{\mu}\f^{\al})^2-V(|\f^{a}|^2),
\end{eqnarray}
with $V(|\f^{a}|^2)$ is a $U(1)$-invariant smooth potential. \re{n1n} is the gauge decoupling limit of \re{n1}.

The scalar field Ansatz factorising the $(\theta,t)$-dependence 
of $\f^{a}$ is
\begin{eqnarray}
\label{hn}
\varphi=\f^{1}-i \f^{2},~~\f^{a}= h(r)\,\hat n^{\al}\ ,
\end{eqnarray}
where $h(r)$ is the (real) scalar amplitude, $n$ is an integer and $\om>0$ is the frequency ($\eta$ is set to 1.)
The fact
that the $(t, \theta)$-dependence of $\f$  occurs in the above form 
implies that the energy-momentum tensor of the model is $t, \theta$ independent.
However, its components will 
 depend on both $n$ and $\om$.
%
One possible expression of the potential $V(|\f^{\al}|^2)$
which allows for spinning vortices with finite mass is given by (\ref{pot1}).

Making the replacements \re{gaugedec} in the reduced Lagrangian of the gauged system \re{lagstt1}
we have the Lagrangian of the 
\begin{eqnarray}
\label{n2}
r^{-1}L =  \lambda_1 \left( h'^2+(\frac{n^2}{r^2}-\om^2)h^2 \right)+ c_3h^2 + c_2h^4 + c_1h^6. 
\end{eqnarray}
The angular momentum density of a spinning  vortex is given by
\begin{eqnarray}
\label{calJ2n}
 {\cal J}= \lambda_1 n\om h^2.
\end{eqnarray}

In contrast to the case of a Higgs (or a Goldstone) field discussed in in Section {\bf 2}, 
the scalar field vanishes asymptotically, $h(r)\to 0$,
such that this model does not possess any topological features.
However, a conserved Noether charge $Q$ is still associated with the complex scalar field $\vf$, 
since the Lagrange density is invariant under
the global phase transformation
$\vf \to \vf e^{i\alpha}$ .
 leading to the conserved current
 \begin{eqnarray}
\label{scalar-current}
j^{\mu}=- \vf^* \partial^{\mu}\vf-\vf \partial^{\mu}\vf^*,~~\nabla_\mu j^{\mu}=0.
 \end{eqnarray}
The corresponding conserved charge $Q$ is the integral of $j^0$.
One can easily see  that  
the following relation holds:
 \begin{eqnarray}
\label{rel1}
J=n\, Q,
\end{eqnarray}
such that  angular momentum  is quantised.
In view of this relation, the spinning vortices
can be thought of as corresponding to minima of energy with fixed angular
momentum.

Spinning solutions of the model (\ref{n1n}) were discussed by several authors,
starting with the early work \cite{Kim:1992mm} (see also \cite{Volkov:2002aj}).
They are the lower dimensional counterparts of the better known Q-balls in 
$(3+1)$-dimensions \cite{Radu:2008pp} and share all their basic properties.
Treating $\om,n$ and the parameters in the potential $U$
as input variables, the
$Q$-vortices
exist only in a certain frequency range, 
$\om_{\rm min} < \om < \om_{\rm max}$.
The limiting behaviour of the spinning solutions at the limits of the $\om$-interval
is rather intricate, and has not been discussed yet in a systematic
way  in the literature. 
It appears that 
both $E$ and $Q$
increase without bounds at the limits of the $\om$-interval.
Also, these configurations do not  always possess a static limit, with $J>0$.

At a critical value of the frequency,
both mass-energy and angular momentum of the solutions assume their minimal value,
from where they monotonically rise towards both limiting values
of the frequency.
Considering the mass of the solutions as a function of the Noether charge $Q$,
there are thus two branches, merging and ending
at the minimal charge and mass. 
The solutions are expected to be stable along the lower branch,
when their mass is smaller than the mass of $Q$ free bosons.

\begin{figure}[h!]
\centering
\includegraphics[height=3in]{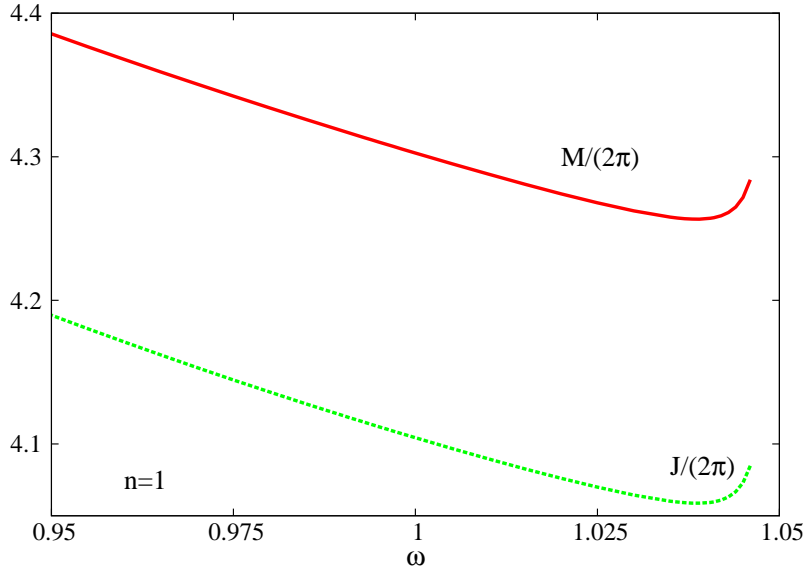}  
\caption{Energy $E$ and angular momentum $J$ are shown as a function of field frequency 
for the typical global (non-topological) vortices. 
}
\label{fig-Qvort}
\end{figure} 

Some results of the numerical integration for $n=1$
are shown in Figure \ref{fig-Qvort}
(note that similar results are found for $n>1$).
 
 \subsection{Gauge decoupled $p=2$ Abelian Higgs model: Spin?}
 
The gauge decoupling limit of the $p=2$ AH model is the $p=2$ Goldstone model 
(see Ref.~\cite{Tchrakian:2010ar} and references therein) whose static Hamiltonian is
\be
\label{ham0G2}
{\cal H}^{(2)}_0=4\la_2(\vep^{\al\beta}\pa_{\mu}\f^{\al}\pa_{\nu}\f^{\beta})^2+24\la_1\,(\eta^2-|\f^{\al}|^2)^2|\pa_{\mu}\f^{\beta}|^2+18\la_0\,(\eta^2-|\f^{\al}|^2)^4 , 
\ee
which in the static axially symmetric limit yields the energy density functional
\bea
\label{hamsttdcp2}
r^{-1}H^{(2)}&=&2\la_2\eta^4\left[r^{-2}\left([(1-h^2)n]'\right)^2\right]\nonumber\\
&&+24\la_1\eta^6(1-h^2)^2\left[\left(h'^2+\frac{n^2h^2}{r^2}\right)\right]+18\la_0\eta^8(1-h^2)^4 , 
\eea
which supports radially symmetric Goldstone vortices.

Our aim here is to consider the axially symmetric stationary Lagrangian corresponding to the static system \re{ham0G2}. In particular, we need the one dimensional reduced Lagrangian of this system
which can be obtained directly from \re{lagstt2} by applying the replacements \re{gaugedec} to it, yielding
\bea
\label{lagstydcp2}
r^{-1}L^{(2)}&=&-2\la_2\eta^4\left[r^{-2}\left([(1-h^2)n]'\right)^2-\left(\omega(1-h^2)'\right)^2\right]\nonumber\\
&&-24\la_1\eta^6(1-h^2)^2\left[\left(h'^2+\frac{n^2h^2}{r^2}\right)-\omega^2\,h^2\right]-18\la_0\eta^8(1-h^2)^4\,.
\eea

The question is, does this system support spin? In constrast to the $p=1$ non-topological vortices considered above,
it is not obvious if this system can sustain nonzero angular momentum. 

First, it is easy to verify the existence of static configurations
in the $\om=0$ limit. However, the numerical accuracy deteriorates very fast with increasing $\om$
without an obvious reason for that behaviour.
As to what the source of this apparent pathology is, may be gleaned by noting that
the quadratic kinetic term in \re{lagstydcp2} is non-standard, and there is no $\mu^2-\om^2$ type coefficient of $h^2$ (for small $h$). Moreover, it is impossible to alter the (reduced) potential $(1-h^2)^4$ by hand, such
as to produce the desired type of coefficient without violating the energy lower bound. This likely indicates the $absence$ of spin for these pure $p=2$ vortices.


\subsection{Stationary ungauged Skyrmion: gauge decoupling limit}
We have seen from our work in Section {\bf 4.2} that the angular momentum of the $U(1)$ gauged Skyrme model vanished in the absence of the Chern-Simons term. (See \re{J1s} when $\ka=0$.)
Thus in the gauge decoupling limit when the Chern-Simons term is absent and $\ka=0$ effectively, one might expect the angular momentum to vanish. Remarkably, this is not the case and
is well known from the work of Ref.~\cite{Piette:1994mh}, that in this limit the angular momentum is supported.

While this fact is known, here we nonetheless verify it for completeness, in concert with
the other two gauge decoupling limits given in Sections {\bf A.1} and {\bf A.2}.

Consider the Lagrangian \re{MaxSk} in the gauge decoupled limit, namely the usual Skyrme model
\be
\label{Lst}
{\cal L}_{\rm Skyrme}=-\frac{1}{8}\la_3|\pa_{[\mu}\ta^a\pa_{\nu]}\ta^b|^2+\frac12\la_1\,|\pa_{\mu}\ta^a|^2-\frac12\la_0\,V[\ta^3]\,,
\ee
and the component of the stress tensor relevant to
the calculation of the angular momentum,
\bea
\label{stressoo}
T_{i0}&=&\frac12\la_3(\pa_{[i}\ta^a\pa_{j]}\ta^b)(\pa_{[j}\ta^a\pa_{0]}\ta^b)
-\la_1\pa_{i}\ta^a\pa_{0}\ta^a\,.
\eea
The stationary Ansatz is adapted from the static Ansatz \re{fa} by replacing the unit vector $n^{\al}$ with $\hat n^{\al}$ given by \re{nhat}. The result is
\be
\label{tiobs}
T_{i0}=\frac{n\,\om}{r}P\left[\la_1P+2\la_3(P'+Q')\right] , 
\ee
leading to the volume integral for the angular momentum
\be
\label{angmomintbs}
J =2\pi\,n\,\om\int\,P\left[\la_1P+2\la_3(P'+Q')\right]rdr , 
\ee
which does not vanish if $\om\neq 0$.
%

\begin{small}

\end{small}

\end{document}